\documentclass[10pt,final,a4paper,twocolumn,twoside,journal,romanappendices]{IEEEtran}
\usepackage{graphicx}
\usepackage{cite}
\usepackage{epstopdf}
\usepackage{epsfig}
\usepackage{amsmath,amssymb,amsfonts,amstext,amsbsy,amsopn,dsfont}
\usepackage{cases}
\usepackage{stfloats}
\usepackage{bigints}
\usepackage{sublabel}
\usepackage{array}

\newtheorem{theorem}{\textbf{Theorem}}

\newtheorem{lemma}{\textbf{Lemma}}

\newtheorem{corollary}{\textbf{Corollary}}
\newcommand{\defn}{\triangleq}
\newcommand{\dif}{\textmd{d}}

\begin{document}

\title{Coverage-Rate Tradeoff Analysis in mmWave Heterogeneous Cellular Networks}


\author{Chun-Hung Liu, \IEEEmembership{Senior Member, IEEE}
%	\thanks{Manuscript received June 28, 2018; revised September 29, 2018; accepted November 11, 2018. Part of this paper was presented at IEEE international Conference on Communications, May 2017 \cite{CHLDCLJRYJC17}.   The work of C.-H. Liu was supported by Mississippi State University under Grant ORED 253551-060702. C.-H. Liu is with the Department of Electrical and Computer Engineering, Mississippi State University, Mississippi State, MS 39762, USA. (e-mail: chliu@ece.msstate.edu).}
	}	

%\markboth{IEEE Transactions on Communications}{Liu: Coverage-Rate Tradeoff Analysis in mmWave Heterogeneous Cellular Networks}

\maketitle

\begin{abstract}
In this paper, we first introduce a generalized modeling and analysis framework to explore the fundamental interactions between user association, coverage probability and link rate in a millimeter wave (mmWave) heterogeneous cellular network (HetNet) in which there are multiple tiers of the ultra-high-frequency (UHF) macrocell and small cell base stations (BSs) and a single tier of mmWave small cell BSs. A generalized user association scheme that can cover many path-loss-based user association schemes is proposed and its related probabilistic properties that facilitate the derivations of the coverage probability and link rate are derived. The derived general expressions of the coverage and link rate not only shed light on how to design user association functions in order to maximize the coverage and link rate but also show that it is impossible to devise a user association scheme that maximizes the coverage and link rate at the same time. Namely, there exists a fundamental tradeoff between the coverage and link rate in mmWave HetNets with distinct bandwidths in the UHF and mmWave bands while a user is associating with a BS. We characterize the coverage-optimal and rate-optimal user association schemes and numerically validate their performances and show the coverage-rate tradeoff problem.
\end{abstract}

\begin{IEEEkeywords}
User association, coverage, link rate, millimeter wave, heterogeneous network, stochastic geometry.	
\end{IEEEkeywords}

\section{Introduction}
\IEEEPARstart{D}{ue to} the proliferation of wireless smart handsets and devices, cellular data traffic is expected to tremendously grow to satisfy customers' huge and different link rate demands in different networking services. To make cellular networks jump over the high-link-rate hurdle due to limited licensed spectrum, densely deploying millimeter wave (mmWave) small cells is a promising approach to alleviate the spectrum crunch problem in the next generation (5G) cellular network. However, mmWave signals suffer high path and penetration losses that significantly weaken the transmission performance of mmWave BSs, especially in an urban area where there are a lot of blockages that severely impede the propagations of mmWave signals \cite{SRTSREE14,TSRGRMMKSSS15,KZLZJMMD15}. The inherent characteristics of mmWave signals bring up a new challenge for the design and deployment of mmWave-based cellular networks. A typical challenge is how to efficiently deploy mmWave (small cell) base stations (BSs) in a blockage environment so that most users can connect to mmWave BSs and enjoy extremely high link rate due to a large available bandwidth in the mmWave band.  In the future architecture of a cellular network, a heterogeneous cellular network (HetNet) is expected to consist of ultra-high-frequency (UHF) macrocells and small cell BSs and mmWave small cell BSs. Such a mmWave HetNet, if compared with the UHF/mmWave stand-alone cellular networks, is anticipated to achieve higher coverage since UHF signals have much lesser penetration and path losses than mmWave signals and has higher link rate since both UHF and mmWave spectra are available.
 
\subsection{Prior Work and Motivations}

In a mmWave HetNet, there arise a few interesting and fundamental problems that are worth investigating. For example, how to efficiently and economically deploy the UHF and mmWave BSs so that they can jointly provide sufficient coverage and data rate in an urban area. As we already knew, mmWave BSs may severely suffer a ``coverage hole'' issue, thereby making (indoor) users isolate from all mmWave BSs due to the weak penetration capability of mmWave signals. Deploying the UHF BSs is able to fill the coverage holes of the mmWave BSs so that the entire network coverage improves. Nevertheless, densely deploying UHF BSs may cause link rate reduction once users tend to associate with the UHF BSs that have a much smaller bandwidth than the mmWave BSs. Hence, a new user association scheme that is able to exploit the advantage of the large bandwidth of the mmWave BSs as well as fill the coverage holes is needed for this mmWave HetNet \cite{DLLWYCMEKKW15}. In addition, how to do traffic offload/loading between the BSs in two different frequency bands is also a paramount problem that needs to be completely studied. Thus, a good modeling and analysis framework needs to be built in order to evaluate the transmission performance (such as coverage and link rate) in mmWave HetNets.
 
The works on the modeling and analysis of a multi-tier HetNet where the UHF and mmWave BSs coexist are still not studied well. Many of the existing works focus on the modeling and performance analysis of cellular networks that merely operate in the mmWave band (typically see \cite{TBRWH15,SSMNKJGA15,XYJZKBL16,AKGAAJGARH16,CWHMW16,MDR15}). Reference \cite{TBRWH15}, for example, studied the coverage and rate problems in a single-tier mmWave cellular network. The approximated analytical results of the coverage probability and rate are obtained by using a simple nearest BS association scheme and neglecting shadowing effects in all channels. In \cite{SSMNKJGA15}, the rate problem was studied in a single-tier mmWave network with a limited self-backhaul resource. The analytical results in the work were obtained based on some simple assumptions, such as, the BSs that are away from users by some critical distance all have non-line-of-sight (NLOS) channels and users receive noise-limited mmWave signals.  Reference \cite{XYJZKBL16} studied how the coverage in a dense mmWave network was affected by the sizes of the antenna arrays and showed that there exists a huge coverage discrepancy between the simplified and actual antenna patterns. In \cite{AKGJGARWH16}, the coverage and rate problems were studied in a single-tier mmWave network where two primary and secondary operators share the same mmWave band, whereas how they were impacted by different user association schemes in this kind of spectrum-sharing operation was not investigated. Although the coverage problem of a multi-tier mmWave cellular network with BS cooperation was studied in \cite{DMNDDT16}, the network completely consists of heterogeneous mmWave BSs and no UHF BSs are in the network. Accordingly, we cannot see how the coverage is jointly affected by the cooperation between UHF BSs and mmWave BSs in this work.  Although a recent work in \cite{HEMNKFBJGA16} indeed studied the coverage problem in a mmWave HetNet consisting of UHF BSs and mmWave small cell BSs, it only focused on the analysis in the uplink and downlink decoupling scenario. It did not investigate how the coverage is contributed by different BSs in different tiers in the non-decoupling scenario and how different user association schemes affect the coverage and rate performances.

\subsection{Contributions}
In the aforementioned prior works, the fundamental interplays between user association, coverage and link rate are not studied at all so that we barely have a clear understanding of the achievable coverage and rate limits even for a single-tier mmWave cellular network. In this work, we aim to thoroughly and generally study the fundamental interactions between user association, coverage and link rate in a mmWave HetNet that is comprised of multiple tiers of the UHF BSs and a single tier of mmWave BSs. The BSs in each tier are of the same type and performance and they are assumed to form an independent Poisson point process (PPP). For analytical tractability, this mmWave HetNet is assumed to be in a blockage environment where all blockages also form an independent PPP with a certain intensity. Under this network model, we first study the statistical fundamental properties of the generalized user association (GUA) scheme that characterizes the general line-of-sight (LOS) and NLOS channel models, blockage effects and user association parameters, and these properties can be easily applied to any specific path loss model and user association scheme. Our network and channel models are much more general than those currently proposed in the literature. This is our first contribution.   

Afterwards we define the signal-to-interference plus noise ratio (SINR) of a user that characterizes the SINRs in the UHF and mmWave bands and use it to define the coverage probability. With the aid of the derived probabilistic properties of the GUA scheme, we derive an accurate expression of the coverage probability for the GUA scheme, which is our second contribution. This derived coverage probability contains a few salient features that are addressed as follows. It clearly indicates how the BSs in each tier contribute the coverage probability so that we are able to know how to efficiently deploy BSs so as to improve the coverage probability. It also shows how multiple antennas, LOS and NLOS channel modeling parameters, user association parameters and blockage intensity influence the coverage probability, and most importantly it indicates that the NLOS BSs and LOS BSs can be viewed as several independent \textit{inhomogeneous} PPPs due to blockages. Moreover, it is so general that it can be applied to some particular/simpler cases, such as the interference-limited case in the UHF band and/or noise-limited case in the mmWave band.

Our third contribution is to find the accurate expression of the link rate of a user. Such a link rate expression contains some identical parameters and functions that also exist in the derived expression of the coverage probability so that it essentially inherits the aforementioned salient features of the derived expression of the coverage probability. It clearly shows how the BSs in each tier contribute their link rate when the GUA scheme is adopted and whether the mmWave BSs could dominate the overall link rate due to their huge bandwidth. These derived general expressions of the coverage and link rate shed light on a fundamental tradeoff problem between the coverage and link rate in a mmWave HetNet, i.e., maximizing the coverage and link rate at the same time by using the same user association scheme is impossible as long as the frequency bands of UHF and mmWave are different. Accordingly, we characterize the coverage-optimal user association scheme and the the rate-optimal user association scheme and show that these two schemes have to use different user association functions. The coverage probabilities and link rates for these two schemes are also accurately found and numerical results are provided to validate their accuracy and the coverage-rate tradeoff problem. This is our fourth contribution.

\subsection{Paper Organization}
The rest of this paper is organized as follows. In Section \ref{Sec:SystemModel}, we first specify the network model in which multi-tier UHF BSs and single-tier mmWave BSs coexist and we then introduce the GUA scheme and channel models for the UHF and mmWave BSs. Section \ref{Sec:CoverageRate} elaborates on how to analyze the coverage and rate for the GUA scheme. In Section \ref{Sec:Tradeoff}, optimal user association schemes that maximize the coverage and rate are studied and how the schemes induce the coverage-rate tradeoff problem is expounded. In Section \ref{Sec:Simulation}, some numerical results are provided to validate the derived analytical results as well as verify the coverage-rate tradeoff finding. Finally, Section \ref{Sec:Conclusion} summarizes our findings and observations. 

\section{System Model and Assumptions}\label{Sec:SystemModel}
%\subsection{Heterogeneous Network Model}
In this paper, we consider an $M$-tier  HetNet over an infinitely large plane in which all BSs in any particular tier that have the same type and performance form an independent PPP with a certain intensity\footnote{Please note that considering independent homogeneous PPPs in an infinitely large plane facilitates the following analyses since the elegant statistical properties of the homogeneous PPPs over an infinitely large plane can make the analyses much tractable.}. To characterize the situation that traditional UHF/microwave BSs  and mmWave (small cell) BSs coexist in this HetNet, we assume the first $M-1$ tiers consist of the UHF macrocell and small cell BSs whereas the $M$th tier consists of the mmWave small cell BSs. For the BSs in the $m$th tier, they can be written as a homogeneous PPP of intensity $\lambda_m$ given by
\begin{align}
	\Phi_m\defn\left\{X_{m,i}\in\mathbb{R}^2: i\in\mathbb{N}_+\right\},
\end{align} 
where $m\in\mathcal{M}\defn\{1,2,\dots,M\}$ and $X_{m,i}$ denotes BS $i$ in the $m$th tier and its location. 

Without of loss of generality, we assume there is a typical user located at the origin and our following location-dependent expressions and analyses are based on this typical user\footnote{According to the  Slivnyak theorem, the statistical properties observed by the typical user located at the origin are the same as those observed by users in any other locations in the network \cite{DSWKJM13}\cite{MH12}.}. Also, we consider the mmWave HetNet is in an urban area where the centers of all blockages (such as buildings, towers, houses, obstacles, etc.) are also assumed to jointly form an independent PPP of intensity $\beta$ for analytical tractability. With considering the blockage effects on the transmission channels between a BS and its serving user, a channel is LOS or NLOS depending on whether or not the channel is visually blocked between the BS and its user. LOS and NLOS channels induced by urban blockages have a very distinct impact on the transmitted signal powers, especially the mmWave signal powers.  

\begin{figure}[!t]
	\centering
	\includegraphics[width=3.25in,height=3in]{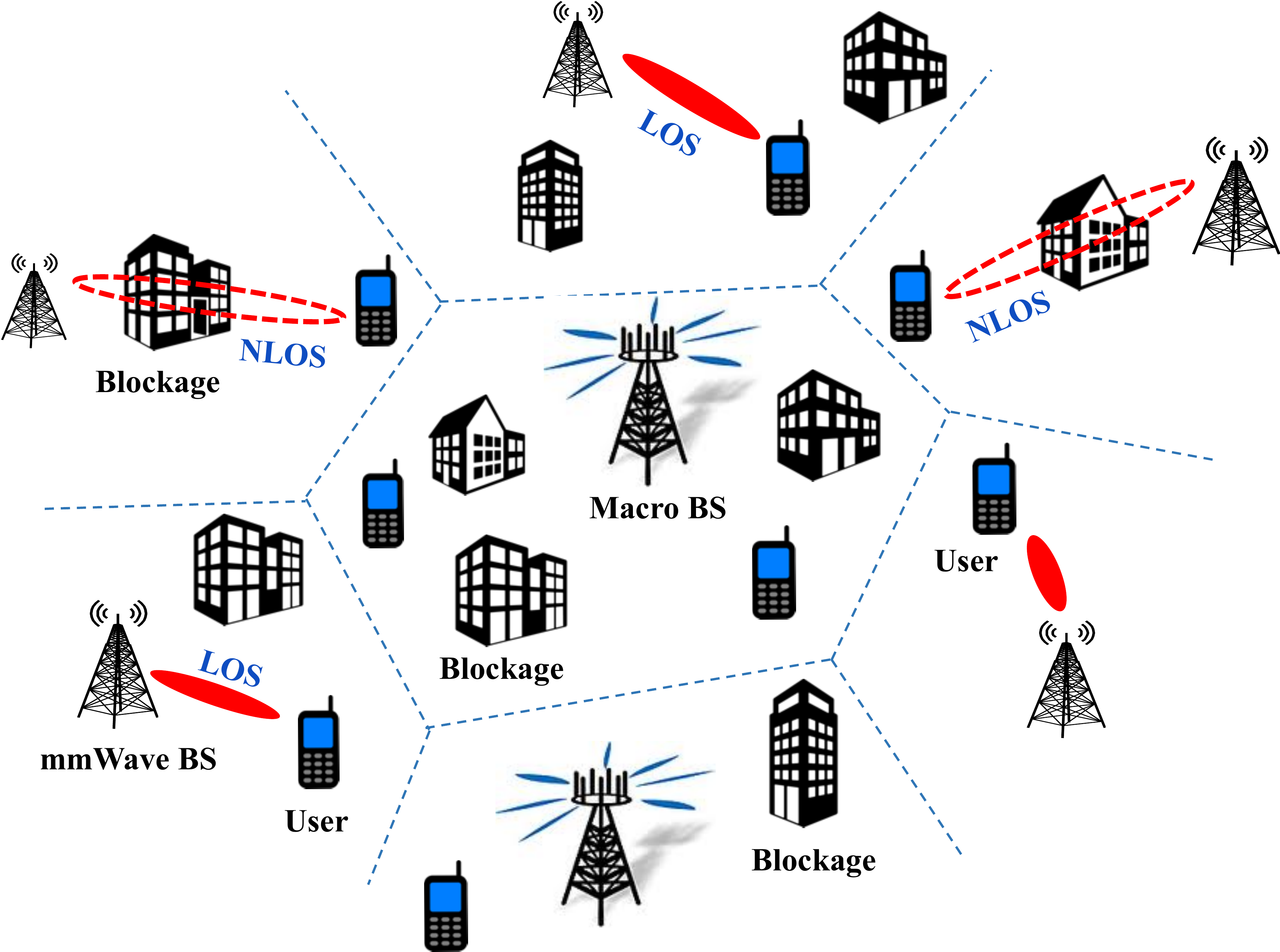}
	\caption{A schematic example of a two-tier mmWave HetNet in an urban area. In this HetNet, all the UHF macro BSs form an independent PPP $\Psi_1$ of intensity $\lambda_1$ and all the mmWave small cell BSs form another independent PPP $\Phi_2$ of intensity $\lambda_2$. In addition, the centers of all the blockages also form an independent PPP of intensity $\beta$. Some of the channels from BSs to users are LOS, whereas some are NLOS due to blockages.}
	\label{Fig:SystemModel}
\end{figure}

\begin{table*}
	\centering
	\caption {NOTATION OF MAIN VARIABLES, SYMBOLS AND FUNCTIONS} \label{Tab:Notation}
	\begin{tabular}{ |c|c||c|c|}
			\hline
			Symbol & Meaning  & Symbol & Meaning \\ 
			\hline
			$\Phi_{m}$ & Set of tier-$m$ BSs & $\alpha_{m,i}$ & Path loss exponent of BS $X_{m,i}$ \\
			$\lambda_m$ &  Intensity of $\Phi_m$ & $F_{Z}(x) $ & CDF of Random Variable (RV) Z \\ 
			$P_m$ & Transmit power of the tier-$m$ BSs & $\phi_m$ &  Tier-$m$ association probability \\ $X_{m,i}$ & BS $i$ in the $m$th tier and its location & $T_m$ &  Number of transmit antennas of tier-$m$ BSs \\ $\beta$ &  Intensity of blockages & $p_{cov}$ & Coverage probability\\  $\eta$ & Geometric parameter of blockages & $W_{\mu} (W_{\varepsilon})$ & Bandwidth of UHF (mmWave) BSs\\$\|Y_i-Y_j\|$  & Distance between nodes $Y_i$ and $Y_j$ & $\nu_{\mu} (\nu_{\varepsilon})$ & Intercept of UHF (mmWave) BSs \\ $L_{m,i}(\cdot)$ & Path loss function of BS $X_{m,i}$ & $g^{-1}(\cdot)$ & Inverse function of $g(\cdot)$\\ $X_*$ & BS associated by the typical user & $g_i\circ g_j$ & Composition of functions $g_{i}$ and $g_j$\\ $G_{m,i}$ &  Shadowing gain for BS $X_{m,i}$ & $\gamma_*$ & SINR at the typical user\\ $H_{m,i}$ &  Channel gain for BS $X_{m,i}$ &
			$I_{*,\mu}$& Interference from all UHF BSs \\ $\Psi_{m,i}(\cdot)$ & User association function of BS $X_{m,i}$ & $I_{*,\varepsilon}$ & Interference from all mmWave BSs  \\$\omega_{m}$ & Association bias of $\Psi_{m,i}$ in \eqref{Eqn:UserAssFunCovOpt} & $C_m (C_*)$& Downlink rate of tier-$m$ BSs (BS $X_*$) \\$\overline{V} (\widetilde{V})$ & Variable (Function) $V$ for LOS (NLOS) channels & $\mathds{1}(\mathcal{E})$ & Indicator function of event $\mathcal{E}$ \\
			\hline
	\end{tabular}
\end{table*}

A schematic example of a two-tier mmWave HetNet is shown in Fig. \ref{Fig:SystemModel} and the notations of main variables, symbols and functions used in this paper are listed in Table \ref{Tab:Notation}. In the following subsection, we will present a generalized user association (GUA) scheme that characterizes the power of the user association signals  (usually called primary synchronization signals in an LTE system)  periodically broadcast by BSs  

\subsection{Generalized User Association (GUA) and Related Statistics}
In this mmWave HetNet, users associate with their serving BS by using the following GUA scheme that is based on the location of the typical user: 
\begin{align}\label{Eqn:GenUserAssoScheme}
X_*&\defn \arg\Psi_*(\|X_*\|) \nonumber\\
&=\arg\sup_{m,i:X_{m,i}\in\Phi} \Psi_{m,i}(\|X_{m,i}\|),
\end{align}
where $X_*\in\Phi\defn \bigcup_{m=1}^{M}\Phi_m$ denotes the BS associated with the typical user, $\|X_i-X_j\|$ denotes the Euclidean distance between BSs $X_i$ and $X_j$ for $i\neq j$, $\Psi_{m,i}:\mathbb{R}_+\rightarrow\mathbb{R}_+$ is called the user association function of BS $X_{m,i}$ and $\Psi_*\in\{\Psi_{m,i}: m\in\mathcal{M}, i\in\mathbb{N}_+\}$ is the user association function of BS $X_*$ and $\Psi_*(\|X_*\|)\defn \sup_{m,i:X_{m,i}\in\Phi}\Psi_{m,i}(\|X_{m,i}\|)$. Since all BSs are in an urban environment, whether their channels to their users are LOS or NLOS is seriously affected by the blockages (especially for the mmWave BSs whose LOS and NLOS channels behave very distinctly.) so that we propose the following user association function $\Psi_{m,i}$ that characterizes the LOS/NLOS channel status of BS $X_{m,i}$:
\begin{align}\label{Eqn:UserAssoFunction}
	\Psi_{m,i}(\|X_{m,i}\|)\defn& \ell(\|X_{m,i}\|)\overline{\Psi}_{m,i}(\|X_{m,i}\|)\nonumber\\
	&+[1-\ell(\|X_{m,i}\|)]\widetilde{\Psi}_{m,i}(\|X_{m,i}\|),
\end{align}
where $\ell(r)\in\{0,1\}$ denotes a Bernoulli random variable (RV) that is one if there is no blockage within distance $r$ and zero otherwise, and $\overline{\Psi}_{m,i}: \mathbb{R}_+\rightarrow\mathbb{R}_+$ ($\widetilde{\Psi}_{m,i}: \mathbb{R}_+\rightarrow\mathbb{R}_+$) is called LOS (NLOS) user association function of BS $X_{m,i}$ having a LOS (NLOS) channel. Both $\widetilde{\Psi}_{m,i}(\cdot)$ and $\overline{\Psi}_{m,i}(\cdot)$ are a (random) monotonic decreasing function since they are supposed to characterize the path loss gain of the user association signals periodically broadcast by BS $X_{m,i}$.  All $\widetilde{\Psi}_{m,i}(\cdot)$'s and $\overline{\Psi}_{m,i}(\cdot)$'s are i.i.d. for the same tier index $m$. 

The distribution of $\Psi_*(\|X_*\|)$ is important for our subsequent analyses, which is shown in the following theorem.
\begin{theorem}\label{Thm:CDFAssBS}
	Suppose the GUA scheme in \eqref{Eqn:GenUserAssoScheme} is adopted. If $\widetilde{\Psi}_{m,i}(\cdot)$ and $\overline{\Psi}_{m,i}(\cdot)$ both are bijective, the cumulative density function (CDF) of $\Psi_*(\|X_*\|)$ is shown as
	\begin{align}\label{Eqn:CDFAssGainBS}
		F_{\Psi_*(\|X_*\|)}(x) =\exp\left(-\pi\sum_{m=1}^{M}\lambda_m\mathsf{A}_m(x)\right),
	\end{align}
	where $\mathsf{A}_m(x)$ is defined as
	\begin{align}\label{Eqn:ExpressionAkm1}
		\mathsf{A}_m(x)\defn 2\mathbb{E}\left[\int_{\widetilde{\Psi}_m^{-1}(x)}^{\overline{\Psi}_m^{-1}(x)} t e^{-\eta\beta t}\dif t\right]+\mathbb{E}\left[\left(\widetilde{\Psi}^{-1}_m(x)\right)^2\right],
	\end{align} 
	where $g^{-1}(\cdot)$ denotes the inverse of real-valued function $g(\cdot)$. The tier-$m$ association probability that users associate with a tier-$m$ BS is
	\begin{align}\label{Eqn:GenAssProbTierM}
		\phi_m=&2\pi\lambda_m\times\nonumber\\
		&\mathbb{E}_{\Psi^{\dagger}_m}\left\{\bigintsss_{0}^{\infty} \exp\left(-\pi\sum_{k=1}^{M}\lambda_k\mathsf{A}_k\circ\Psi^{\dagger}_m(x)\right)x\dif x\right\},
	\end{align}
	where composition function $\mathsf{A}_k\circ\Psi^{\dagger}_m(x)$ is defined as
	\begin{align}\label{Eqn:ExpressionAkm3}
		\mathsf{A}_k\circ\Psi^{\dagger}_m(x)=&\mathbb{E}\bigg[\int_{\widetilde{\Psi}^{-1}_k\circ\widetilde{\Psi}_m^{\dagger}(x)}^{\overline{\Psi}^{-1}_k\circ\overline{\Psi}_m^{\dagger}(x)}2te^{-\eta\beta t}\dif t+\nonumber\\
		&\left(\widetilde{\Psi}^{-1}_k\circ\widetilde{\Psi}^{\dagger}_m(x)\right)^2\bigg|\Psi^{\dagger}_m(x)\bigg]
	\end{align}
	in which $g_k\circ g_m(x)=g_k(g_m(x))$ is the composition of functions $g_k(\cdot)$ and $g_m(\cdot)$ and functions $g^{\dagger}_m(\cdot)$ and $g_m(\cdot)$ are i.i.d. for all $m\in\mathcal{M}$ if they are random. 
\end{theorem}
\begin{IEEEproof}
	See Appendix \ref{App:ProofCDFAssBS}.
\end{IEEEproof}

The CDF in \eqref{Eqn:CDFAssGainBS} is so general that it not only works for any invertible user associate functions but also contains the impacts of LOS and NLOS channels. For example, $\mathsf{A}_m(x)$ in \eqref{Eqn:ExpressionAkm1} can be explicitly found if $\overline{\Psi}_m(x)$ and $\widetilde{\Psi}_m(x)$ are designed as an invertible power-law path loss function of $x$, as shown in the following subsection. Furthermore, we can realize $\mathbb{E}\left[(\widetilde{\Psi}^{-1}_m(x))^2\right] \leq \mathsf{A}_m(x)\leq\mathbb{E}\left[(\overline{\Psi}^{-1}_m(x))^2\right] $ because $\lim_{\beta\rightarrow\infty}\mathsf{A}_m(x)=\mathbb{E}\left[(\widetilde{\Psi}^{-1}_m(x))^2\right] $ (i.e., an infinitely large blockage intensity makes all channels become NLOS) and $\lim_{\beta\rightarrow 0}\mathsf{A}_m(x)=\mathbb{E}\left[(\overline{\Psi}^{-1}_m(x))^2\right]$ (i.e., all channels are LOS because of no blockages). As such, assuming all channels are NLOS after some distance away from the typical user (e.g., the LOS ball model proposed in \cite{SSMNKJGA15}), which is the popular modeling assumption made in the prior related works, may significantly impact the accuracy of the analytical results especially when the network is dense. 

\subsection{Path Loss and Channel Gain Models for UHF and mmWave BSs}\label{SubSec:ChannelModels}
\textbf{\textit{Path Loss Models}}. The signals of all BSs undergo path loss before they arrive at their serving user. In this paper, we consider the following path loss function $L_{m,i}(\cdot)$ between BS $X_{m,i}$ and the typical user:
\begin{align}\label{Eqn:PathLossModel}
L_{m,i}(\|X_{m,i}\|)\defn \nu_m\|X_{m,i}\|^{\alpha_{m,i}} 
\end{align}
where $\|X_{m,i}\|$ denotes the Euclidean distance between BS $X_{m,i}$ and the typical user, $\nu_m\defn\nu_{\mu}\mathds{1}(X_{m,i}\notin\Phi_M)+\nu_{\varepsilon}\mathds{1}(X_{m,i}\in\Phi_M)$ in which $\mathds{1}(\mathcal{E})$ is the indicator function that is equal to one if event $\mathcal{E}$ is true and zero otherwise, $\nu_{\mu}$ and $\nu_{\varepsilon}$ denote  the intercepts\footnote{Here we assume that $\nu_{\mu}$ and $\nu_{\varepsilon}$ both contain the closed-in free-space path-loss so that in this paper we still call $\alpha_{m,i}$ path loss exponent, which is defined slightly different from the terminologies used in the previous works on the mmWave channel models \cite{TSRRHRDJM14,TSRGRMMKSSS15}.} of the UHF signals and the mmWave signals, respectively, $\alpha_{m,i}$ is called the \textit{path-loss exponent} that characterizes the LOS path loss exponent $\overline{\alpha}$ and the NLOS path loss exponent $\widetilde{\alpha}$ of BS $X_{m,i}$, and it is written as\footnote{In practice, the path-loss exponents of the BSs in different tiers are more likely to be different, i.e., $\overline{\alpha}$ and $\widetilde{\alpha}$ should be different when it is used in different tiers. For the analytical tractability in this paper, however, we still assume that the path loss exponents of the BSs in different tiers are the same.}
\begin{align}\label{Eqn:DefnPathloss}
\alpha_{m,i}\defn \ell(\|X_{m,i}\|)\overline{\alpha}+[1-\ell(\|X_{m,i}\|)]\widetilde{\alpha},
\end{align} 
where $\overline{\alpha}$ ($\widetilde{\alpha}$) is the LOS (NLOS) path-loss exponent of a BS. Note that we assume $\overline{\alpha}<\widetilde{\alpha}$ since LOS channels usually should have lesser path loss than NLOS channels. According to \cite{TBRVRWH14}, we know $\mathbb{P}[\ell(r)=1]=e^{-\eta\beta r}$ where $\eta$ is a geometric parameter regarding to the mean perimeter of blockages\footnote{For example, $\eta$ is equal to $\frac{1}{\pi}\times $ the mean perimeter of a rectangular blockage \cite{TBRVRWH14}.}. %The feature of the pathloss $\alpha$ in \eqref{Eqn:DefnPathloss} is that it characterizes how the BS intensities and the blockage intensity impact $\alpha$. For instance, $\alpha$ reduces to $\widetilde{\alpha}$ as $\beta$ goes to infinity (i.e., all channels become NLOS due to too many blockages), and $\alpha$ becomes $\overline{\alpha}$ as $\beta$ goes to zero (i.e., all channels are LOS due to no blockages).  

The results in Theorem \ref{Thm:CDFAssBS} regarding the GUA scheme can be simplified to make themselves much implementable based on the path loss model in \eqref{Eqn:PathLossModel}. Namely, we can consider the user association function in \eqref{Eqn:UserAssoFunction} pertaining to the path loss of the BSs. That is, we can assume
\begin{align}\label{Eqn:PowerLawUserAssFun}
	\Psi_{m,i}(\|X_{m,i}\|)\defn \frac{\psi_{m,i}}{L_{m,i}(\|X_{m,i}\|)},
\end{align}
where $\psi_{m,i}\defn \ell(\|X_{m,i}\|)\overline{\psi}_{m,i}+[1-\ell(\|X_{m,i}\|)]\widetilde{\psi}_{m,i}$. Parameter $\overline{\psi}_{m,i}$ ($\widetilde{\psi}_{m,i}$) can be viewed as the random path loss bias when BS $X_{m,i}$ has a LOS (NLOS) channel\footnote{Note that $\overline{\psi}_{m,i}$ and $\widetilde{\psi}_{m,i}$ are usually designed to characterize the random channel gain such as fading and/or shadowing in user association signals that are periodically emitted by \textit{omni-directional} antennas of BS $X_{m,i}$.}. Based on the user association function in \eqref{Eqn:PowerLawUserAssFun}, we simplify the results in Theorem \ref{Thm:CDFAssBS} in the following corollary. 
\begin{corollary}\label{Cor:CDFAssBS}
	If the user association function of BS $X_{m,i}$ in \eqref{Eqn:PowerLawUserAssFun} is adopted with $L_{m,i}(\cdot)$ in \eqref{Eqn:PathLossModel}, the results in \eqref{Eqn:ExpressionAkm1} and \eqref{Eqn:ExpressionAkm3} reduce to the following:
	\begin{align}\label{Eqn:PowerLawExpressionAkm1}
		\mathsf{A}_m(x) =\mathbb{E}\left[\int_{(\frac{\widetilde{\psi}_m}{x})^{1/\widetilde{\alpha}}}^{(\frac{\overline{\psi}_m}{x})^{1/\overline{\alpha}}} 2te^{-\eta\beta t}\dif t\right]+\mathbb{E}\left[\left(\frac{\widetilde{\psi}_m}{x}\right)^{\frac{2}{\widetilde{\alpha}}}\right]
	\end{align}
	and
	\begin{align}\label{Eqn:PowerLawExpressionAkm2}
		\mathsf{A}_k\circ\Psi^{\dagger}_m(x) = & x^2\bigg(\mathbb{E}\left[\int_{\left(\frac{\widetilde{\psi}_k}{\widetilde{\psi}^{\dagger}_m}\right)^{1/\widetilde{\alpha}}}^{\left(\frac{\overline{\psi}_k}{\overline{\psi}^{\dagger}_m}\right)^{1/\overline{\alpha}}}2te^{-\eta\beta x t}\dif t\bigg| \Psi^{\dagger}_m(x)\right]\nonumber\\
		&+\mathbb{E}\left[\left(\frac{\widetilde{\psi}_k}{\widetilde{\psi}^{\dagger}_m}\right)^{2/\widetilde{\alpha}}\bigg| \widetilde{\psi}^{\dagger}_m\right]\bigg).
	\end{align}
	Furthermore, if all $\overline{\psi}_{m,i}$'s and $\widetilde{\psi}_{m,i}$'s are deterministic, the CDF of $\|X_*\|$ is explicitly found as
	\begin{align}
		F_{\|X_*\|}(x)=1-\sum_{m=1}^{M} \phi_m \exp\left(-\pi\sum_{k=1}^{M}\lambda_k\mathsf{A}_k\circ\Psi_m(x)\right),
		%\frac{2}{(\eta\beta)^2}\left[\gamma\left(2,\eta\beta\left(\frac{\overline{\psi}_m}{x}\right)^{\frac{1}{\overline{\alpha}}}\right)-\gamma\left(2,\eta\beta\left(\frac{\widetilde{\psi}_m}{x}\right)^{\frac{1}{\widetilde{\alpha}}}\right)\right]+\left(\frac{\widetilde{\psi}_m}{x}\right)^{\frac{2}{\widetilde{\alpha}}}.
	\end{align}
	where $\phi_m$ and $\mathsf{A}_k\circ\Psi_m(x)$ are
	\begin{align}
		\phi_m &= 2\pi\lambda_m \int_{0}^{\infty} \exp\left(-\pi\sum_{k=1}^{M}\lambda_k\mathsf{A}_k\circ\Psi_m(x)\right) x\dif x,\label{Eqn:AssProbTierm} \end{align}
	\begin{align}	
		\mathsf{A}_k\circ\Psi_m(x) =&x^2\bigg[\int_{\left(\frac{\widetilde{\psi}_k}{\widetilde{\psi}_m}\right)^{1/\widetilde{\alpha}}}^{\left(\frac{\overline{\psi}_k}{\overline{\psi}_m}\right)^{1/\overline{\alpha}}}2te^{-\eta\beta xt}\dif t\nonumber\\
		&+\left(\frac{\widetilde{\psi}_k}{\widetilde{\psi}_m}\right)^{2/\widetilde{\alpha}}\bigg].\label{Eqn:DetPowerLawExpressionAkm}
	\end{align}
\end{corollary}
\begin{IEEEproof}
	Since $\overline{\Psi}_{m,i}(x)=\overline{\psi}_{m,i}x^{-\overline{\alpha}}$ and $\widetilde{\Psi}_{m,i}(x)=\widetilde{\psi}_{m,i}x^{-\widetilde{\alpha}}$, their inverse functions are given by $\overline{\Psi}^{-1}_{m,i}(x)=(\overline{\psi}_{m,i}/x)^{1/\overline{\alpha}}$ and $\widetilde{\Psi}^{-1}_{m,i}(x)=(\widetilde{\psi}_{m,i}/x)^{1/\widetilde{\alpha}}$, respectively. Substituting these explicit results of $\overline{\Psi}_{m,i}(x)$, $\overline{\Psi}^{-1}_{m,i}(x)$, $\widetilde{\Psi}_{m,i}(x)$ and $\widetilde{\Psi}^{-1}_{m,i}(x)$ into \eqref{Eqn:ExpressionAkm1} and \eqref{Eqn:ExpressionAkm3} results in \eqref{Eqn:PowerLawExpressionAkm1} and \eqref{Eqn:PowerLawExpressionAkm2}. In addition, for all deterministic monotonic decreasing $\overline{\Psi}_{m,i}$'s and $\widetilde{\Psi}_{m,i}$'s $F_{\|X_*\|}(x)$ can be alternatively expressed as
	\begin{align*}
		F_{\|X_*\|}(x)&=1-\sum_{m=1}^{M}\phi_m F_{\Psi_m(\|X_*\|)}\left(\Psi_m(x)\right)\nonumber\\
		&=1-\sum_{m=1}^{M}\phi_m\exp\left(-\pi\sum_{k=1}^{M}\lambda_k\mathsf{A}_k\circ\Psi_m(x)\right),
	\end{align*} 
	where $\mathsf{A}_k\circ\Psi_m(x)=\mathsf{A}_k\circ\Psi^{\dagger}_m(x)$ because $\Psi_m(x)=\Psi^{\dagger}_m(x)$ in the deterministic case. Hence, $\mathsf{A}_k\circ\Psi_m(x)$ in \eqref{Eqn:DetPowerLawExpressionAkm} can be readily acquired from \eqref{Eqn:PowerLawExpressionAkm2} in the deterministic case.
\end{IEEEproof}
\noindent With the results in Corollary \ref{Cor:CDFAssBS}, we can find the statistical properties of the biased power-law path loss of the associated BS and the association probability of each tier for any power-law path-loss-based association policies, such as nearest BS association, maximum mean received power association, green cell association, etc. \cite{CHLLCW16,CHLKLF16,CHL16}. 

%We will expound how to use Corollary \ref{Cor:CDFAssBS} to analyze the coverage probability and the achievable rate of a user for our newly proposed rate-aware user association scheme in Section \ref{Sec:CoverageThroughput}.  

\textbf{\textit{Channel Fading and Shadowing Gain Models}}. Suppose the BSs in the $m$th tier are equipped with $T_m$ transmit antennas and all users are equipped with a single antenna, i.e., we have a multiple-input-single-output (MISO) channel from a BS to a user\footnote{To make the analyses much tractable in this paper, users are only considered to be equipped with a single antenna so that all analyses in the downlink are performed based on the MISO channel model. }. According to reference \cite{MKYLMSSSSR14}, the fading gain vector of a mmWave MISO channel can be properly represented by a clustered channel model consisting of small-scale fading and angle-of-departure (AoD)-based transmit array gain vectors. Also, we assume that all BSs have a uniform linear array and are able to perfectly align their beam with the AoD of their array in order to maximize their antenna array gain. 

When BS $X_*$ equipped with $T_*\in\{T_1,\ldots, T_M\}$ transmit antennas performs transmit beamforming to the typical user, the MISO fading channel gain from it to the user is expressed as 
\begin{align}\label{Eqn:MISOChannelModelSerBS}
H_*\defn  [h_{*,\mu}\mathds{1}(X_*\notin\Phi_M)+h_{*,\varepsilon}\mathds{1}(X_*\in\Phi_M)]G_*,
\end{align} 
where $h_{*,\mu}$ ($h_{*,\varepsilon}$) denotes the small-scale fading gain in the UHF (mmWave) band and $G_*=\overline{G}_*\ell(\|X_*\|)+\widetilde{G}_*[1-\ell(\|X_*\|)]$ is the large-scale shadowing gain in which $\overline{G}_*$ and $\widetilde{G}_*$ denote the LOS and NLOS shadowing gains, respectively. We assume each BS has the channel state information of its users so that $h_{*,\mu}\sim\chi^2_{2T_*}$ is a Chi-squared RV with $2T_*$ degrees of freedom and $h_{*,\varepsilon}\sim T_*\exp(1)$ is an exponential RV with mean $T_*$ and variance $T^2_*$ due to transmit beamforming performed by BS $X_*$. Moreover, if if BS $X_*\in\Phi_m$, $\overline{G}_*\sim\mathcal{\ln N}(0,\overline{\rho}^2_m)$ and $\widetilde{G}_*\sim\mathcal{\ln N}(0,\widetilde{\rho}^2_m)$ are log-normal RVs that are zero mean and have variances $\overline{\rho}^2_m$ and $\widetilde{\rho}^2_m$, respectively.  Note that we usually have $\widetilde{\rho}^2_m>\overline{\rho}^2_m$ for all $m\in\mathcal{M}$ since NLOS channels usually suffer a larger shadowing variation in LOS channels based on many previous measurement results \cite{TSRGRMMKSSS15,TSRRHRDJM14}. Similarly, the interference channel gain from BS $X_{m,i}$ to the typical user can be written as
\begin{align}\label{Eqn:MISOChannelModelIntfBS}
H_{m,i}=h_{m,i}G_{m,i},
\end{align}
where $h_{m,i}\sim\mathrm{exp}(1)$\footnote{Unlike the fading gain of the communication channel in \eqref{Eqn:MISOChannelModelSerBS}, the fading gain in the interference channels does not contain the effect of the multi-antenna diversity because the interfering BSs cannot do transmit beamforming to the typical user \cite{SWJGA10,PXCHLJGA13}.} and $G_{m,i}=\overline{G}_{m,i}\ell(\|X_{m,i}\|)+\widetilde{G}_{m,i}[1-\ell(\|X_{m,i}\|)]$ in which $\overline{G}_{m,i}\sim\ln\mathcal{N}(0,\overline{\rho}^2_m)$ and $\widetilde{G}_{m,i}\sim\ln\mathcal{N}(0,\widetilde{\rho}^2_m)$. Note that all $\widetilde{G}_{m,i}$'s and $\overline{G}_{m,i}$'s are independent for all $i\in\mathbb{N}_+$ and $m\in\mathcal{M}$,  and they are i.i.d. for the same tier index $m$.  All $h_{m,i}$'s are i.i.d for all $i\in\mathbb{N}_+$ and $m\in\mathcal{M}$.

\subsection{The SINR Model for the UHF and mmWave Bands}
According to the GUA scheme with the user association function designed in \eqref{Eqn:PowerLawUserAssFun}, the general expression of the signal-to-interference plus noise power ratio (SINR) of the typical user associating with BS $X_*$ can be written as
\begin{align}\label{Eqn:GeneralSINR}
\mathrm{SINR}_* = \frac{H_*P_*}{I'_*L_*(\|X_*\|)},
\end{align}
where $P_*\in\{P_1,\ldots,P_M\}$ is the power of BS $X_*$ and $P_m$ is the power of the tier-$m$ BSs,  interference $I'_*\defn(I_{*,\mu}+\sigma^2_{\mu})\mathds{1}(X_*\notin\Phi_M)+(I_{*,\varepsilon}+\sigma^2_{\varepsilon})\mathds{1}(X_*\in\Phi_M)$ in which $\sigma^2_{\mu}$ ($\sigma^2_{\varepsilon}$) denotes the noise power in the UHF (mmWave) band, $I_{*,\mu}$ (interference in the UHF band) and $I_{*,\varepsilon}$ (interference in the mmWave band) are given by
\begin{align}
I_{*,\mu} &= \sum_{m,i:X_{m,i}\in\bigcup_{m=1}^{M-1}\Phi_m\setminus\{X_*\}} \frac{P_mH_{m,i}}{L_{m,i}(\|X_{m,i}\|)},\\
I_{*,\varepsilon}&= \sum_{M,i:X_{M,i}\in\Phi_M\setminus\{X_*\}} \frac{P_MH_{M,i}}{L_{M,i}(\|X_{M,i}\|)},
\end{align}
respectively.
 
The SINR model in \eqref{Eqn:GeneralSINR} may be simplified to another low-complexity model by considering the practical signal propagation characteristics in the UHF and mmWave bands. In the UHF band, the interference usually dominates the received signal power so that the UHF BSs are interference-limited in general, whereas channels in the mmWave band would significantly suffer a non-negligible noise power due to their large bandwidth\cite{SRTSREE14,SSMNKJGA15}. In this case,  the SINR in \eqref{Eqn:GeneralSINR} can be accurately approximated by $\gamma_*(X_*)$ defined as
\begin{align}\label{Eqn:ApporxSINR}
\gamma_*(X_*) \defn \frac{P_*H_*}{I_*L_*(\|X_*\|)}  
\end{align}
by assuming $I_{*,\mu}\gg \sigma^2_{\mu}$ almost surely and thus $I'_*\approx I_*\defn I_{*,\mu}\mathds{1}(X_*\notin\Phi_M)+(I_{*,\varepsilon}+\sigma^2_{\varepsilon})\mathds{1}(X_*\in\Phi_M)$. Namely, we can consider an SIR model in the UHF band and an SINR model in the mmWave band. In the following sections, we will use the SINR model in \eqref{Eqn:ApporxSINR} to characterize the coverage probability and link rate.

\section{Coverage and Rate Analysis for Generalized User Association}\label{Sec:CoverageRate}
In this section, we focus on the analysis of the coverage probability and link rate in the downlink when the GUA scheme with the biased path-loss-based user association \eqref{Eqn:PowerLawUserAssFun} is adopted:
\begin{align}\label{Eqn:UserAssFunCovOpt}
\Psi_{m,i}(\|X_{m,i}\|)=\frac{\omega_mG_{m,i}}{L_{m,i}(\|X_{m,i}\|)},
\end{align}
where $\omega_{m}G_{m,i}$ is equal to $\psi_{m,i}$ in \eqref{Eqn:PowerLawUserAssFun} and $\omega_{m}$ is a constant bias for the tier-$m$ BSs. We first define the coverage probability based on the SINR defined in \eqref{Eqn:ApporxSINR} and then derive the general approximated expression of the coverage probability with the GUA scheme in \eqref{Eqn:GenUserAssoScheme}.  We then analyze the achievable link rate of a user and explicitly find its approximated accurate expression. 
%According to the derived expressions of the coverage and rate, we can characterize the fundamental tradeoff problem between coverage and link rate in a mmWave HetNet, which will be elaborated in Section \ref{Sec:Tradeoff}.

\subsection{Coverage Probability Analysis}
Suppose the SINR threshold for success decoding at each user is $\theta$. By using the SINR model in \eqref{Eqn:ApporxSINR}, the (downlink) coverage probability of a user in the mmWave HetNet is defined as
\begin{align}
p_{cov}(\theta)&\defn\mathbb{P}\left[\gamma_*\geq\theta\right]\nonumber\\
&=\sum_{m=1}^{M}\phi_m\mathbb{P}[\gamma_*\geq \theta|X_*\in\Phi_m],
\end{align}
where the tier-$m$ association probability $\phi_m=\mathbb{P}[X_*\in\Phi_m]$ is already found in \eqref{Eqn:AssProbTierm}. Using $\gamma_*$ in \eqref{Eqn:ApporxSINR} leads to $p_{cov}(\theta)$ explicitly given  by
\begin{align}\label{Eqn:DefnCoverageProb}
p_{cov}(\theta) =\sum_{m=1}^{M} \phi_m \mathbb{P}\left[\frac{P_mH_m}{I_*L_*(\|X_*\|)}\geq\theta\right].
\end{align}
Note that $p_{cov}(\theta)$ in \eqref{Eqn:DefnCoverageProb} is a spacial average and ergodic result which can be shown by using the Campbell and Slinvyak theorems. Since the distribution of $L_*(\|X_*\|)$ depends on how the user association function is designed, $p_{cov}(\theta)$ highly depends on the user association scheme. The user association signals emitted from the BSs usually undergo small-scale fading and large-scale shadowing whereas only the fading component in the signals are usually able to be averaged out at users. Hence, for the user association function \eqref{Eqn:PowerLawUserAssFun} characterizing the shadowing gain, the coverage probability in \eqref{Eqn:DefnCoverageProb} can be explicitly found as shown in the following theorem.
\begin{theorem}\label{Thm:CoverageProbBias}
If the user association function in \eqref{Eqn:UserAssFunCovOpt}  is adopted, then the coverage probability in \eqref{Eqn:DefnCoverageProb} can be explicitly approximated as
\begin{align}\label{Eqn:CoverageProbBias}
p_{cov}(\theta) \approx& \sum_{m=1}^{M-1}\phi_m\left(\sum_{n=0}^{T_m-1}\frac{(-\theta)^n}{n!}\frac{\dif^n}{\dif\theta^n}\mathsf{B}_m(\theta)\right) \nonumber\\
&+ \phi_M \mathsf{B}_M(\theta),
\end{align}
where $\phi_m$ is given by
\begin{align}
\phi_m =& 2\pi\lambda_m\times\nonumber\\
&\mathbb{E}_{\Psi^{\dagger}_m}\left\{\bigintsss_{0}^{\infty} \exp\left(-\pi\sum_{k=1}^{M}\lambda_k\mathsf{A}_k\circ\Psi^{\dagger}_m(x)\right)x\dif x\right\},\label{Eqn:AssProbTiermBias}
\end{align}
\begin{align}
\mathsf{A}_k\circ\Psi^{\dagger}_m(x) =& x^2\bigg\{\mathbb{E}\left[\int_{\left(\frac{\widetilde{G}_k\nu_m\omega_k}{\widetilde{G}^{\dagger}_m\nu_k\omega_m}\right)^{1/\widetilde{\alpha}}}^{\left(\frac{\overline{G}_k\nu_m\omega_k}{\overline{G}^{\dagger}_m\nu_k\omega_m}\right)^{1/\overline{\alpha}}}2t e^{-\eta\beta xt}\dif t\bigg|G_m^{\dagger}\right]\nonumber\\
& + e^{4\frac{\widetilde{\rho}^2_k}{\widetilde{\alpha}^2}}\left(\frac{\omega_k}{\nu_k}\widetilde{G}^{\dagger}_m\right)^{2/\widetilde{\alpha}}\bigg\}\label{Eqn:AkPsimBias}
\end{align}
with i.i.d. RVs ($\overline{G}_m$, $\overline{G}^{\dagger}_m$) and i.i.d. RVs ($\widetilde{G}_m$, $\widetilde{G}^{\dagger}_m$), $\mathsf{B}_m(\theta)$ for $m\neq M$ is given in \eqref{Eqn:BmfunBias}
in which $\Lambda(q,s,r)\defn\sum_{k=1}^{M}\Lambda_k(q,s,r)$ 
\begin{figure*}[!h]
	\begin{align}
	\mathsf{B}_m(\theta)\approx\bigintsss_{0}^{\infty}\frac{2\pi x\Lambda(0,0,x) \dif x}{\exp\left(2\pi \left[\sum_{k=1}^{M-1} \int^{\infty}_x\Lambda_k\left(\frac{r}{x},\frac{P_m}{\theta\omega_m},r\right) r\dif r +\int_{0}^{x}\Lambda(0,0,r)r\dif r\right]\right)} , \label{Eqn:BmfunBias}
	\end{align}
\end{figure*}
and $\Lambda_k(\cdot,\cdot,\cdot)$ is defined as\footnote{Note that $\Lambda_m(r)$ in \eqref{Eqn:TiermIntensityBias} will reduce to $\overline{\lambda}_me^{-\eta\beta r}$ as $\widetilde{\alpha}\rightarrow\infty$. This corresponds to the case that the penetration and path losses are so large that the BSs with NLOS channels essentially cannot be detected by any users and usually mmWave BSs are in this kind of situation.}
\begin{align}
%\Lambda(r)\defn & \sum_{m=1}^{M}\Lambda_m(r),\label{Eqn:TotalIntensityBias}\\
\Lambda_k\left(q,s,r\right) \defn& \frac{ \lambda_k\omega_k^{\frac{2}{\overline{\alpha}}}e^{4\frac{\overline{\rho}^2_k}{\overline{\alpha}^2}}e^{-\eta\beta r}}{q^{\overline{\alpha}}\frac{s\omega_k}{ P_k}+1}\nonumber\\
&+\frac{\lambda_k\omega^{\frac{2}{\widetilde{\alpha}}}_ke^{4\frac{\widetilde{\rho}^2_k}{\widetilde{\alpha}^2}}(1-e^{-\eta\beta r})}{q^{\widetilde{\alpha}}\frac{s\omega_k}{  P_k}+1},\label{Eqn:TiermIntensityBias}
\end{align}
and $B_M(\theta)$ is given in \eqref{Eqn:BMfunBias}.
\begin{figure*}[!h]
	\begin{align}
	\mathsf{B}_M(\theta)\approx&\bigintsss_{0}^{\infty}\frac{2\pi x\Lambda(0,0,x)\exp\left(-\frac{\theta \omega_M\nu_M\sigma^2_{\varepsilon}}{T_MP_M}\left[e^{-\eta\beta x}\left(x^{\overline{\alpha}}-x^{\widetilde{\alpha}}\right)+x^{\widetilde{\alpha}}\right]\right)}{\exp\left(2\pi  \left[\int_{0}^{x}\Lambda(0,0,r)r\dif r+\int^{\infty}_x\Lambda_M\left(\frac{r}{x},\frac{P_M}{\theta\omega_M},r\right) r\dif r\right]\right) }\dif x.\label{Eqn:BMfunBias}
	\end{align}
	\hrulefill
\end{figure*}
\end{theorem}
\begin{IEEEproof}
See Appendix \ref{App:ProofCoverageProbBias}.
\end{IEEEproof}

Theorem \ref{Thm:CoverageProbBias} reveals a few important implications. First, the coverage probability in \eqref{Eqn:CoverageProbBias} reflects how the coverage is contributed by the BSs across two different frequency spectra so that it gives us a much clear understanding of how to deploy mmWave BSs and UHF BSs and do traffic loading/offloading between different tiers by changing biases $\omega_m$'s in order to effectively improve the coverage under a given blockage intensity. Second, the physical meanings of $\mathsf{B}_m(\theta)$ in \eqref{Eqn:BmfunBias} and $\mathsf{B}_M(\theta)$ in \eqref{Eqn:BMfunBias} are the coverage probabilities contributed by the tier-$m$ UHF BSs\footnote{We can show that $\mathsf{B}_m(\theta)$ actually can exactly reduce to the coverage probability found in some previous works, such as \cite{HSJYJSXPJGA12}.} and the tier-$M$ mmWave BSs, respectively. As a result, we can obviously see how much the coverage in \eqref{Eqn:CoverageProbBias} is improved by adding more antennas. Third, the coverage probability in a mmWave multi-tier HetNet is certainly better than that in a single-tier mmWave cellular network since users (such as indoor users) still can be covered by the UHF BSs if they are not well covered by the mmWave BSs. We can clearly observe this phenomenon from the asymptotic values of $\mathsf{B}_m(\theta)$ in \eqref{Eqn:BmfunBias} and $\mathsf{B}_M(\theta)$ in \eqref{Eqn:BMfunBias} by letting $\lambda_M$ goes to infinity. Fourth, since the NLOS and LOS channels are assumed to independently suffer different shadowing gains, the coverage probability is characterized by $m$ independent \textit{inhomogeneous} PPPs whose \textit{distance-dependent intensities} are shown in \eqref{Eqn:TiermIntensityBias}, which is the main reason that $\mathsf{B}_m$ in \eqref{Eqn:BmfunBias} and $\mathsf{B}_M$ in \eqref{Eqn:BMfunBias} cannot be further simplified provided the blockage effects need to be generally and exactly characterized in the interference model. 

By considering some particular realistic channel characteristics in the UHF and mmWave bands, the results in Theorem \ref{Thm:CoverageProbBias} would be largely simplified. For example, the transmitted mmWave signals usually suffer fairly large penetration loss as well as noise power, whereas the LOS and NLOS channels in the UHF band can be simply modeled by a unified channel model (e.g., 3GPP adopts a unified channel model for LOS and NLOS channels in an urban area \cite{36.872}.). That is, we can consider $\widetilde{G}_M\rightarrow 0$ and $\widetilde{\alpha}_M\rightarrow\infty $ in the mmWave band to represent huge penetration loss and the mmWave BSs are assumed to have a LOS channel if the distance between them and their user is not greater than $d_L$. Consider  $\overline{\alpha}=\widetilde{\alpha}=\alpha_{\mu}$ and $\overline{G}_m=\widetilde{G}_m=G_m\sim\ln\mathcal{N}(0,\rho^2_m)$ for the all UHF channels. Thus, we have
\begin{align}
\mathsf{A}_k\circ\Psi^{\dagger}_m(x)\approx\begin{cases}x^2 e^{4\frac{\sigma^2_k}{\alpha^2_{\mu}}}\left(\frac{\omega_kG_m}{\nu_k}\right)^{\frac{2}{\alpha_{\mu}}}, k,m\neq M\\
x^2e^{4\frac{\rho^2_m}{\alpha^2_{\mu}}-\eta\beta d_L}\left(\frac{\omega_MG_m}{\nu_M}\right)^{\frac{2}{\alpha_{\mu}}}, k=M, m\neq M\\
x^2e^{4\frac{\rho^2_m}{\alpha^2_{\mu}}-\eta\beta d_L}\left(\frac{\omega_MG_M}{\nu_M}\right)^{\frac{2}{\alpha_{\mu}}}, k=m=M
\end{cases}
\end{align}
and then substitute it into \eqref{Eqn:AssProbTiermBias} to find $\phi_m$ approximately as shown in \eqref{Eqn:AssProbTiermBias2}.
\begin{figure*}[!t]
\begin{align}
\phi_m\approx\frac{\lambda_m(\omega_m/\nu_{\mu})^{\frac{2}{\alpha_\mu}}e^{4\frac{\rho^2_m}{\alpha^2_{\mu}}}\mathds{1}(m\neq M)+\lambda_M(\omega_M/\nu_{\varepsilon})^{\frac{2}{\overline{\alpha}}}e^{4\frac{\rho^2_M}{\alpha^2_{\mu}}-\eta\beta d_L}\mathds{1}(m= M)}{\sum_{k=1}^{M-1}\lambda_k(\omega_k/\nu_k)^{\frac{2}{\alpha_\mu}}e^{4\frac{\rho^2_m}{\alpha^2_{\mu}}}+\lambda_M(\omega_M/\nu_{\varepsilon})^{\frac{2}{\alpha_{\mu}}}e^{4\frac{\rho^2_M}{\alpha^2_{\mu}}-\eta\beta d_L}}.\label{Eqn:AssProbTiermBias2}
\end{align}
\end{figure*}
Also, $\Lambda_m(q,s,r)$ in \eqref{Eqn:TiermIntensityBias} reduces to
\begin{align}
\Lambda_m(q,s,r)\approx& \left(\frac{\lambda_m\omega^{\frac{2}{\alpha_{\mu}}}_me^{4\frac{\rho^2_m}{\alpha_{\mu}^2}}}{q^{\alpha_{\mu}}\frac{s\omega_m}{ P_m}+1}\right)\mathds{1}(m\neq M)\nonumber\\
&+\left(\frac{\lambda_M\omega^{\frac{2}{\alpha_{\mu}}}_Me^{-\eta\beta d_L+4\frac{\rho^2_M}{\alpha^2_{\mu}}}}{q^{\alpha_{\mu}}\frac{s\omega_M}{ P_M}+1}\right)\mathds{1}(m=M), \label{Eqn:TiermIntensityBias2}
\end{align}
$\Lambda=\sum_{m=1}^{M}\Lambda_m$, and $\mathsf{B}_m(\theta)$ in \eqref{Eqn:BmfunBias} and $\mathsf{B}_M(\theta)$ in \eqref{Eqn:BMfunBias} reduce to \eqref{Eqn:BmfunBias2} and \eqref{Eqn:BMfunBias2}respectively.  
\begin{figure*}[!t]
\begin{align}
\mathsf{B}_m(\theta)\approx \left[1+\sum^{M-1}_{k=1}\phi_k\left(\frac{\theta P_k\omega_m}{P_m\omega_k}\right)^{\frac{2}{\alpha_{\mu}}}\left(\frac{2\pi/\alpha_{\mu}}{\sin(2\pi/\alpha_{\mu})}-\int_{0}^{\left(\frac{\theta P_k\omega_m}{P_m\omega_k}\right)^{-\frac{2}{\alpha_{\mu}}}}\frac{\dif t}{1+t^{\frac{\alpha_{\mu}}{2}}}\right)\right]^{-1},\label{Eqn:BmfunBias2}
\end{align}
\begin{align}
\mathsf{B}_M(\theta)\approx\bigintsss_{0}^{\infty}\frac{2\pi x\Lambda(0,0,x) \dif x}{\exp\left(2\pi  \left[\frac{1}{2}x^2\Lambda(0,0,x)+\int^{\infty}_x\Lambda_M\left(\frac{r}{x},\frac{P_M}{\theta\omega_M},r\right)r\dif r\right]+\frac{\theta \omega_M\sigma^2_{\varepsilon}\nu_M}{T_MP_Me^{\eta\beta x}}x^{\alpha_{\mu}}\right)},\label{Eqn:BMfunBias2}
\end{align}
\hrulefill
\end{figure*}
As can be seen, the results in \eqref{Eqn:BmfunBias2} and \eqref{Eqn:BMfunBias2} are significantly simplified by comparing their corresponding results in \eqref{Eqn:BmfunBias} and \eqref{Eqn:BMfunBias}.

\subsection{Link Rate Analysis}
In this subsection, our focus is on the analysis of the link rate of a user for the GUA scheme with $\Psi_{m,i}$ given in \eqref{Eqn:UserAssFunCovOpt}. Let $W_{\mu}$ and $W_{\varepsilon}$ denote the available bandwidths of the UHF and mmWave BSs, respectively\footnote{Note that in general we have $W_{\mu}\gg W_{\varepsilon}$ because the available bandwidth of mmWave BSs is significantly larger than that of UHF BSs.}.  We can define the (achievable) downlink rate of the typical user as
\begin{align}\label{Eqn:DefErgodicLinkThroughput}
C_*\defn W_*\mathbb{E}\left[\ln(1+\gamma_*(X_*))\right],\quad (\text{nats/sec})
\end{align}
where $W_*=W_{\mu}\mathds{1}(X_*\notin\Phi_M)+W_{\varepsilon}\mathds{1}(X_*\in\Phi_M)$ and $\gamma_*(\cdot)$ is defined in \eqref{Eqn:ApporxSINR}. $C_*$ can be further expressed as 
\begin{align}\label{Eqn:DefnErgodicLinkThroughput}
C_*=\sum_{m=1}^{M}\phi_m C_{m},
\end{align}
where $C_{m}$ denotes the link rate of the BSs in the $m$th tier and it can be written as
\begin{align}\label{Eqn:DefnLinkRateInEachTier}
C_m=\begin{cases} W_{\mu} \mathbb{E}\left[\ln\left(1+\frac{P_mH_m}{I_{*,\mu}L_*(\|X_*\|)}\right)\right], & m\neq M \\ W_{\varepsilon}\mathbb{E}\left[\ln\left(1+\frac{P_MH_M}{(I_{*,\varepsilon}+\sigma^2_{\varepsilon})L_*(\|X_*\|)}\right)\right], & m=M  \end{cases}.
\end{align}
In the following theorem, we show the explicit expression of the link rate with the GUA scheme.
\begin{theorem}\label{Thm:ErgodicLinkRate}
If the user association function in \eqref{Eqn:UserAssFunCovOpt} is adopted, the link rate in \eqref{Eqn:DefnErgodicLinkThroughput} can be approximately characterized as
\begin{align}\label{Eqn:ErgodicLinkThroughput}
\hspace{-0.1in}C_* \approx& W_{\mu}\sum_{m=1}^{M-1}\phi_m\bigintsss_{0}^{\infty}\left[1-  \left(1+\frac{sP_m}{\omega_m}\right)^{-T_m}\right]\times\nonumber\\ &\mathsf{B}_m\left(\frac{P_m}{s\omega_m}\right)\frac{\dif s}{s}+W_{\varepsilon}\phi_M\bigintsss_{0}^{\infty} \frac{T_MP_MB_M(\frac{1}{s})}{\omega_M+sT_MP_M} \dif s, 
\end{align}
where $\phi_m$, $\mathsf{B}_m(\cdot)$ and $\mathsf{B}_M(\cdot)$ are already defined in \eqref{Eqn:AssProbTiermBias}, \eqref{Eqn:BmfunBias} and \eqref{Eqn:BMfunBias}, respectively. 
%\begin{align}
%\mathsf{D}_{\varepsilon}(s)\defn&  \int_{0}^{\infty}2\pi x \Lambda(x) \exp\left(-2\pi\left[ \int_{0}^{x}\Lambda(r)r\dif r+\int^{\infty}_xr\Lambda_M(r)\hbar_{\frac{1}{s}}\left(\left(\frac{x}{r}\right)^{\overline{\alpha}},\left(\frac{x}{r}\right)^{\widetilde{\alpha}};\eta\beta r\right) \dif r\right] \right)\nonumber\\
%&-\frac{s \omega_M}{P_M}W_{\varepsilon}\nu_MN_0\left[e^{-\eta\beta x}\left(x^{\overline{\alpha}}-x^{\widetilde{\alpha}}\right)+x^{\widetilde{\alpha}}\right]\bigg) \dif x,
%\end{align}
%$\phi_m$ is already defined in \eqref{Eqn:AssProbTiermBias} for all $m\in\mathcal{M}$.
\end{theorem} 
\begin{IEEEproof}
See Appendix \ref{App:ProofErgodicLinkRate}.
\end{IEEEproof}

The expression of the link rate in \eqref{Eqn:ErgodicLinkThroughput} has a few salient features that are worth being addressed in the following. First of all, it is a very general result that characterizes the LOS and NLOS channels, blockage impact as well as MISO fading in a low-complexity form; to the best of our knowledge, it is never derived in previous works with a PPP-based network model. It also characterizes how $C_*$ changes with the user association biases so that it can be used for several different user association schemes and indicate how to do traffic offloading/loading between tiers in order to significantly improve $C_*$. Moreover, it clearly shows how the BSs in each tier contribute $C_*$ so that we are able to know how to efficiently deploy the BSs in every tier to significantly improve $C_*$. For example, increasing tier-$m$ bias $\omega_M$ (or deploying more mmWave BSs) can make traffic offload to the mmWave tier and it should increase $C_*$ in general since $W_{\varepsilon}$ is extremely larger than $W_{\mu}$. However, in a dense blockage area offloading too much traffic to the mmWave tier may not significantly improve $C_*$ due to the large penetration loss of mmWave signals. 

In addition,  the computational complexity in \eqref{Eqn:ErgodicLinkThroughput} is not high for many practical contexts because $C_*$ in \eqref{Eqn:ErgodicLinkThroughput} is actually expressed in terms of $B_m(\cdot)$ and $B_M(\cdot)$ and thus it can be largely simplified for some practical contexts, as shown in the previous coverage analysis. For example, if we use a unified channel model for all LOS and NLOS channels in the UHF band and consider a huge penetration loss of the channels in the mmWave band, in this scenario $C_*$ in \eqref{Eqn:ErgodicLinkThroughput} can be largely simplified since $B_m\left(\frac{P_m}{s\omega_m}\right)$ and $B_M\left(\frac{1}{s}\right)$ can be found by using the results in \eqref{Eqn:BmfunBias2} and \eqref{Eqn:BMfunBias2}, respectively.  In addition, we should be aware that the coverage probability and link rate are derived based on the user association function in \eqref{Eqn:UserAssFunCovOpt}. Thus, here arises a fundamental question: \textit{ Can we maximize the coverage and the link rate at the same time by using the same user association function in \eqref{Eqn:UserAssFunCovOpt}}? We will study this question in the following section.

\section{Optimal User Association Schemes and Coverage-Rate Tradeoff}\label{Sec:Tradeoff}
In this section, we would like to investigate a fundamental question of how to design the user association function $\Psi_{m,i}(\cdot)$ that is able to maximize the coverage probability and/or link rate in a mmWave HetNet. Namely, we want to find the coverage-optimal association (COA) scheme that maximizes the coverage probability  and the rate-optimal association (ROA) scheme that maximizes the link rate. Also, we also want to explore the fundamental relationship between COA and ROA.   

In the following lemma, we summarize our findings for the COA scheme.
\begin{lemma}\label{Lem:COA}
	If we assign $\omega_{m}=P_m$ in the user association function in \eqref{Eqn:UserAssFunCovOpt} (i.e., $\Psi_{m,i}(\|X_{m,i}\|)=\frac{P_mG_{m,i}}{\|X_{m,i}\|^{\alpha_{m,i}}}$), the GUA scheme with such a user association function is the COA scheme. 
\end{lemma}
\begin{IEEEproof}
	See Appendix \ref{App:ProofCOA}.
\end{IEEEproof}
\noindent Accordingly, the COA scheme is obtained by the GUA scheme with $\Psi_{m,i}=\frac{P_m G_{m,i}}{\|X_{m,i}\|^{\alpha_{m,i}}}$ and  it achieves the maximum coverage probability that can be readily obtained by the result in Theorem \ref{Thm:CoverageProbBias} for $\omega_{m}=P_m$. For the ROA scheme, we summarize our findings in the following lemma.
\begin{lemma}\label{Lem:ROA}
	Let bias $\omega_{m}$ be defined as $\omega_{m}=W_{\mu}\mathds{1}(m\neq M)+W_{\varepsilon}\mathds{1}(m=M)$. If  the GUA scheme with $\Psi_{m,i}(\|X_{m,i}\|) = (\frac{P_mG_{m,i}}{\|X_{m,i}\|^{\alpha_{m,i}}})^{\omega_{m}}$ for all $m\in\mathcal{M}$ and $i\in\mathbb{N}_+$, it is the ROA scheme that maximizes the downlink rate of users. Also, if $W_{\mu}=W_{\varepsilon}$, letting $\Psi_{m,i}(\|X_{m,i}\|) = \frac{P_mG_{m,i}}{\|X_{m,i}\|^{\alpha_{m,i}}}$ also maximizes the downlink rate of users. 
\end{lemma} 
\begin{IEEEproof}
	See Appendix \ref{App:ProofROA}.
\end{IEEEproof}

From Lemma \ref{Lem:ROA}, we learn that ROA and COA are the  exactly same scheme that maximizes the coverage and link rate at the same time when the bandwidths in the UHF and mmWave spectra are equal. Also, Lemmas \ref{Lem:COA} and \ref{Lem:ROA} manifest that\textit{ it is impossible to find a user association scheme that is able to maximize the coverage and link rate at the same time if a network has two distinct radio spectra available. In other words, there always exists a coverage-rate tradeoff problem in a mmWave HetNet that has two distinct bandwidths $W_{\mu}$ and $W_{\varepsilon}$ ($W_{\mu}\ll W_{\varepsilon}$)}. In the following subsections, we will study this tradeoff problem in more detail.

\subsection{Achievable Coverage Probability and Link Rate for the COA Scheme} \label{SubSec:CovProbRateCOA}

Since the COA scheme has a user association function $\Psi_{m,i}(\|X_{m,i}\|) = \frac{P_mG_{m,i}}{\|X_{m,i}\|^{\alpha_{m,i}}}$ for BS $X_{m,i}$ as indicated in Lemma 1, the coverage probability achieved by COA can be directly obtained by substituting $\omega_{m}=P_m$  into \eqref{Eqn:CoverageProbBias} and it is given by
\begin{align}\label{Eqn:MaxCovProb}
p_{cov}(\theta) \approx \sum_{m=1}^{M-1}\phi_m\left(\sum_{n=0}^{T_m-1}\frac{(-\theta)^n}{n!}\frac{\dif^n}{\dif\theta^n}\mathsf{B}_m(\theta)\right) + \phi_M \mathsf{B}_M(\theta),
\end{align}
where $\phi_m$, $\phi_M$, $B_m(\cdot)$ and $B_M(\cdot)$ can be found by substituting $\omega_{m}=P_m$ into \eqref{Eqn:AssProbTiermBias}, \eqref{Eqn:BmfunBias} and \eqref{Eqn:BMfunBias}, respectively. Similarly, the link rate achieved by the COA scheme can be found by substituting $\omega_{m}=P_m$ into \eqref{Eqn:ErgodicLinkThroughput} and it is given by
\begin{align}\label{Eqn:LinkRateCOA}
C_* \approx &  W_{\mu}\sum_{m=1}^{M-1}\phi_m\int_{0^+}^{\infty}\left[1-  (1+s)^{-T_m}\right] \mathsf{B}_m\left(\frac{1}{s}\right)\frac{\dif s }{s}\nonumber\\
&+W_{\varepsilon}\phi_M\int_{0}^{\infty} \frac{T_MB_M(1/s)}{1+sT_M} \dif s.
\end{align}
where $\phi_m$, $\phi_M$, $B_m(\cdot)$ and $B_M(\cdot)$ are the corresponding ones already found in \eqref{Eqn:MaxCovProb}. Note that $C_*$ in \eqref{Eqn:LinkRateCOA} is not the maximum achievable rate in the HetNet with $W_{\varepsilon}\gg W_{\mu}$ based on Lemma \ref{Lem:COA} and the maximum achievable link rate will be introduced in the following subsection. 

\subsection{Achievable Coverage Probability and Link Rate for the ROA Scheme} \label{SubSec:CovProbRateROA} 
\begin{figure*}[!b]
	\hrulefill
	\begin{align}\label{Eqn:GenAssProbTierMROA}
	\phi_m=2\pi\lambda_m\mathbb{E}_{\Psi^{\dagger}_m}\left\{\bigintsss_{0}^{\infty} \exp\left[-\pi\sum_{k=1}^{M}\lambda_k\mathsf{A}_k\left(\frac{(P_mG^{\dagger}_m)^{W_{\mu}}}{x^{\alpha_m}}\right)\right]x\dif x\right\},
	\end{align}
	\begin{align}
	\mathsf{B}_m(\theta)\approx&\frac{2\pi}{W_{\mu}}\bigintsss_{0}^{\infty} \frac{x^{\frac{2}{W_{\mu}}-1}\Lambda(0,0,x^{\frac{1}{W_{\mu}}}) \dif x}{\exp\left(2\pi \left[\sum_{k=1}^{M-1} \int^{\infty}_x\Lambda_k\left(\frac{r}{x},\frac{P_m}{\theta W_{\mu}},r\right)  r\dif r+\int_{0}^{x^{\frac{1}{W_{\mu}}}}\Lambda(0,0,r)r\dif r\right]\right) }, \label{Eqn:BmfunBiasROA}
	\end{align}
	\begin{align}
	\mathsf{B}_M(\theta)\approx\frac{2\pi}{W_{\varepsilon}}\bigintsss_{0}^{\infty} \frac{x^{\frac{2}{W_{\varepsilon}}-1}\Lambda(0,0,x^{\frac{1}{W_{\varepsilon}}})\exp\left(-\frac{\theta W_{\varepsilon}\nu_M\sigma^2_{\varepsilon}}{T_MP_M}\left[e^{-\eta\beta x}\left(x^{\overline{\alpha}}-x^{\widetilde{\alpha}}\right)+x^{\widetilde{\alpha}}\right]\right)\dif x}{\exp\left(2\pi  \left[\int_{0}^{x^{\frac{1}{W_{\varepsilon}}}}\Lambda(0,0,r)\dif r+\int^{\infty}_{x}\Lambda_M\left(\frac{r}{x},\frac{P_M}{\theta W_{\varepsilon}},r\right) r\dif r\right]\right)}.\label{Eqn:BMfunBiasROA}
	\end{align}
\end{figure*}
According to Lemma \ref{Lem:ROA}, the coverage probability achieved by the ROA scheme is shown in the following corollary.
\begin{corollary}\label{Cor:CoverageROA}Consider all users adopt the ROA scheme found in Lemma \ref{Lem:ROA} to associate with their BS.  The coverage probability with the ROA scheme can be found by $p_{cov}(\theta)$ in \eqref{Eqn:CoverageProbBias} with $\phi_m$, $B_m(\cdot)$ and $B_M(\cdot)$ as shown in \eqref{Eqn:GenAssProbTierMROA}, \eqref{Eqn:BmfunBiasROA} and \eqref{Eqn:BMfunBiasROA}, respectively.
\end{corollary}
\begin{IEEEproof}
	 From Lemma \ref{Lem:ROA}, we know $\Psi_{m,i}(x) = (\frac{P_mG_{m,i}}{x^{\alpha_{m,i}}})^{\omega_{m}}$ and thus $\Psi^{-1}_{m,i}(x) = (P_mG_{m,i}x^{-\frac{1}{\omega_{m}}})^{\frac{1}{\alpha_{m,i}}}$. Now replacing $x$ in \eqref{Eqn:AssProbTiermBias}, \eqref{Eqn:BmfunBias} and \eqref{Eqn:BMfunBiasROA} with $x^{\frac{1}{\omega_{m}}}$ yields the results in \eqref{Eqn:GenAssProbTierMROA}, \eqref{Eqn:BmfunBiasROA} and \eqref{Eqn:BMfunBiasROA}. 
\end{IEEEproof}
\noindent Note that the coverage probability achieved in Corollary \ref{Cor:CoverageROA} is always smaller than that in \eqref{Eqn:MaxCovProb} based on Lemma \ref{Lem:COA}. We will numerically verify this point in Section \ref{Sec:Simulation}.

Now consider that all users adopt the ROA scheme found in Lemma 2 to associate with their BS. By substituting According to Theorem \ref{Thm:ErgodicLinkRate}, the link rate achieved by the ROA scheme can be readily found as
\begin{align}\label{Eqn:LinkRateROA}
C_* \approx & W_{\mu}\sum_{m=1}^{M-1}\phi_m\bigintsss_{0}^{\infty}\left[1-  \left(1+\frac{sP_m}{W_{\mu}}\right)^{-T_m}\right]\mathsf{B}_m\left(\frac{P_m}{s W_{\mu}}\right)\nonumber\\ 
&\times\frac{1}{s}\dif s+W_{\varepsilon}\phi_M\int_{0}^{\infty} \frac{T_MP_MB_M(\frac{1}{s})}{W_{\varepsilon}+sT_MP_M} \dif s, 
\end{align}
where $\mathsf{B}_m(P_m/s W_{\mu})$ and $B_M(\frac{1}{s})$ can be found by using \eqref{Eqn:BmfunBiasROA} and \eqref{Eqn:BMfunBiasROA} in Corollary \ref{Cor:CoverageROA}.
%\begin{corollary}\label{Cor:RateROA}
%	Suppose . The link rate achieved by the ROA scheme 
%\end{corollary}
\noindent Also, the link rate achieved by ROA in \eqref{Eqn:LinkRateROA} is always higher than that achieved by COA in \eqref{Eqn:LinkRateCOA} based on Lemma \ref{Lem:ROA}.  We can intuitively explain this in more detail as follows. According to the proof of Lemma \ref{Lem:ROA} in Appendix \ref{App:ProofROA}, in the low SINR regime, ROA makes $\phi_M$ increase so that offloading traffic to the mmWave tier in general should increase $C_*$ since usually the link rate increase in $\phi_MC_{\varepsilon}$ is larger than the link rate loss in $\sum_{m=1}^{M-1}\phi_mC_{\mu,m}$ due to $W_{\mu}\ll W_{\varepsilon}$. In the high SINR regime, users are also more likely offloaded to the mmWave tier owing to $W_{\varepsilon}\gg W_{\mu}$. Therefore, from a rate point of view, making users associate with a mmWave BS in general improves their link rate. We will numerically demonstrate this point in the following subsection. 

\section{Numerical Results}\label{Sec:Simulation}

\begin{table*}[!t]
	\centering
	\caption{Network Parameters for Simulation\cite{TSRRHRDJM14}}\label{Tab:SimPara}
	\begin{tabular}{|c|c|c|}
		\hline Parameter $\setminus$ BS Type& Macrocell & mmWave Picocell (73 GHz)\\ 
		\hline Power $P_m$  & 20 (W) & 1(W)\\ 
		\hline Intensity $\lambda_m$  & $1\times 10^{-6}$(BSs/$m^2$) & (see figures)    \\ 
		\hline Number of Antennas $T_m$ & 4 & 2 \\ 
		\hline Bandwidth $W_{\mu}$,$W_{\varepsilon}$  & $0.1$ GHz & $1$ GHz \\
		\hline SIR Threshold $\theta$ &\multicolumn{2}{c|}{1} \\ 
		\hline  $\ln\overline{G}_{m} \sim\mathcal{N}(0,\overline{\rho}^2_m)$ &  $\mathcal{N}(0,13\,\text{dB})$  & $\mathcal{N}(0,9.6\,\text{dB})$\\ 
		\hline  $\ln\widetilde{G}_{m} \sim\mathcal{N}(0,\widetilde{\rho}^2_m)$ &  $\mathcal{N}(0,13\,\text{dB})$  & $\mathcal{N}(0,15.8\,\text{dB})$\\ 
		\hline Path loss Exponent ($\overline{\alpha},\widetilde{\alpha}$) & (2.1, 3.4) & (2.1, 6.75) \\ 
		\hline Blockage Intensity $\beta$ & \multicolumn{2}{c|}{$5.5\times 10^{-5}$ (blockages/m$^2$)} \\
		\hline 
	\end{tabular} 
\end{table*}

\subsection{Numerical Results of the Coverage Probability}\label{SubSec:SimulationCovProb}
\begin{figure*}[!t]
\centering
	\includegraphics[width=\textwidth]{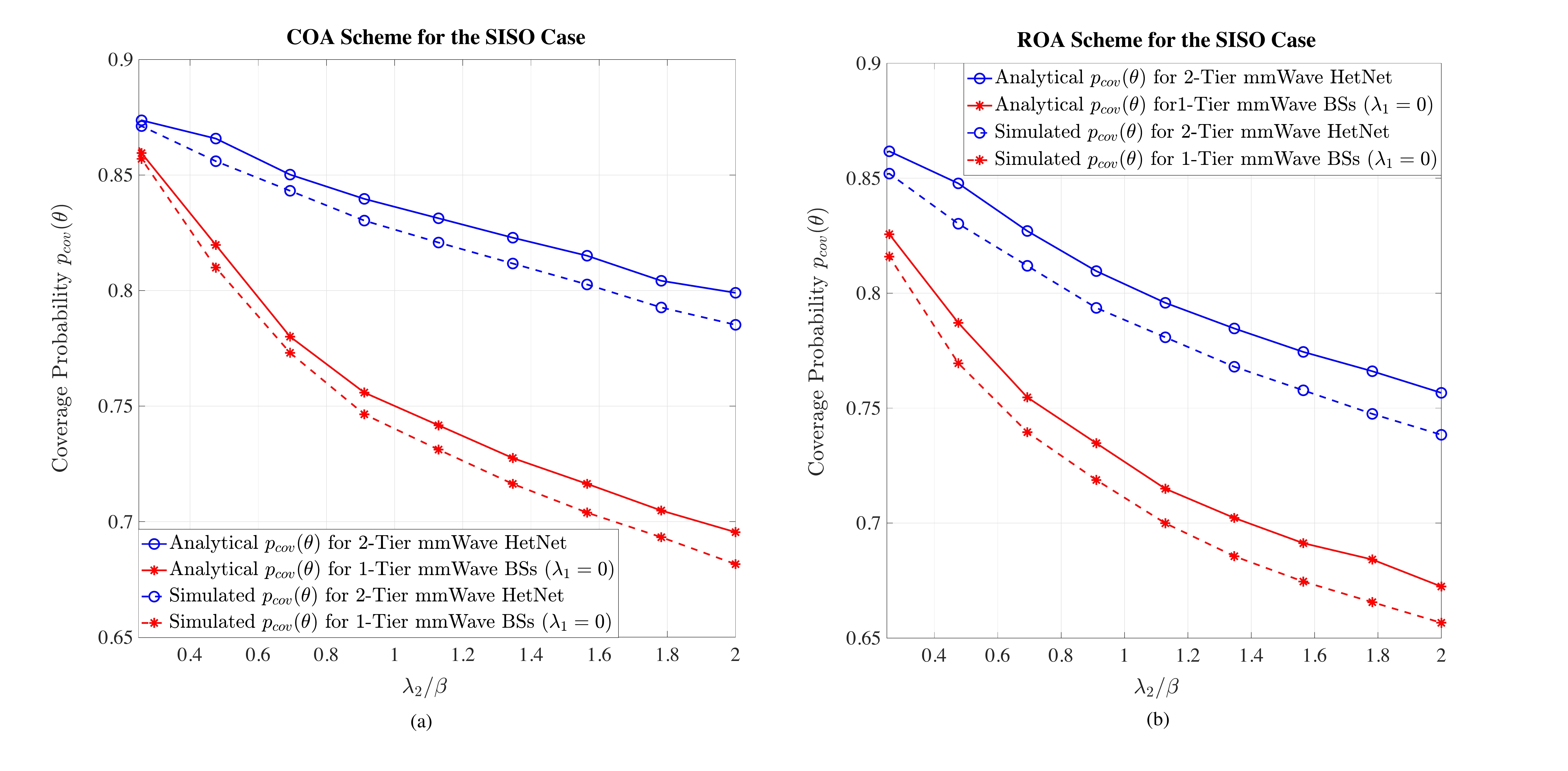}
	\caption{Simulation results of the coverage probability: (a) the COA scheme and (b) the ROA schemes. In this simulation, all BSs and users are equipped with a single antenna, i.e., the SISO case.}
	\label{Fig:CovProbSISO}
\end{figure*}
\begin{figure*}
\centering
	\includegraphics[width=\textwidth]{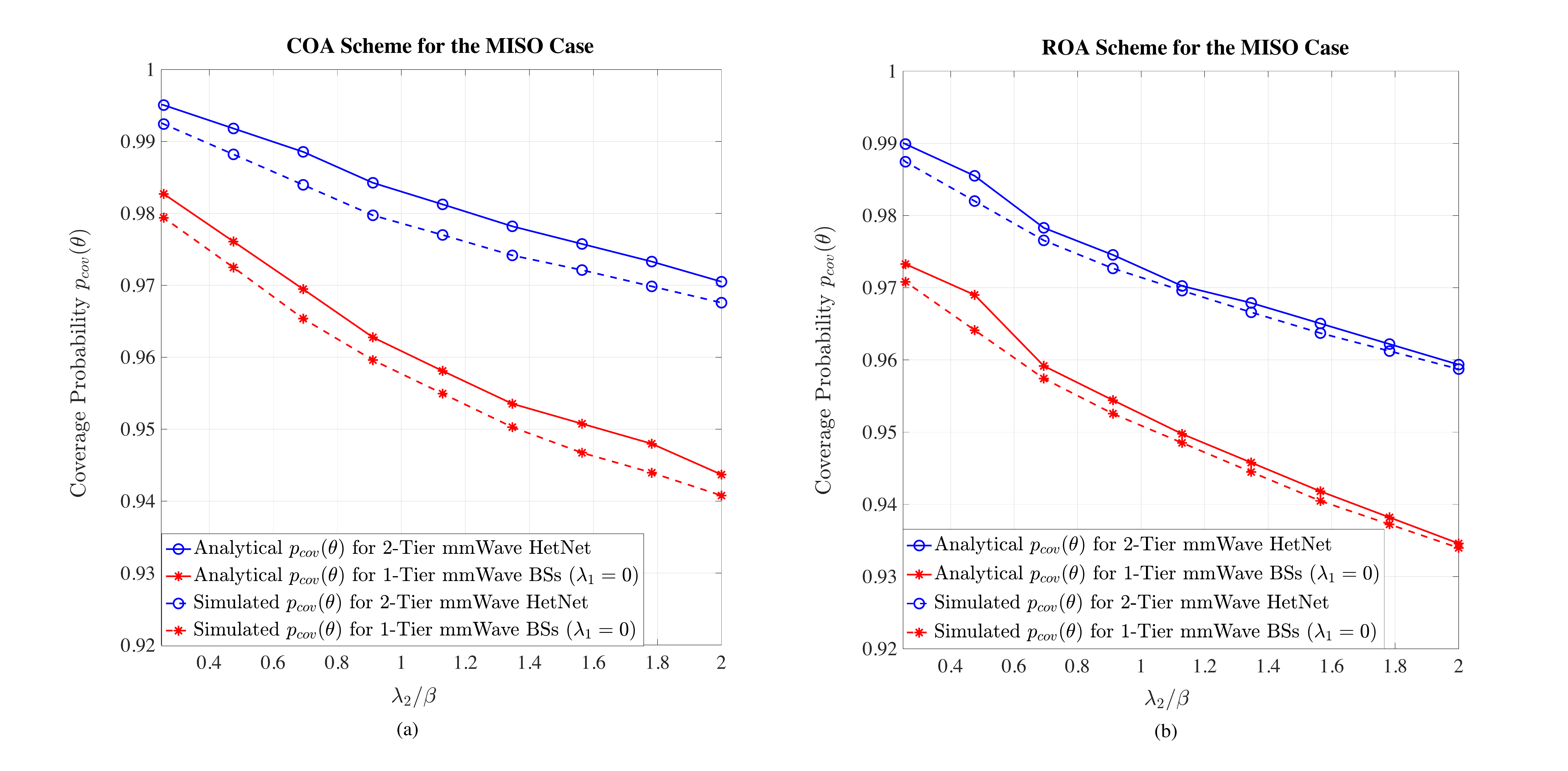}
	\caption{Simulation results of the coverage probability: (a) the COA scheme and (b) the ROA schemes. In this simulation, all BSs are equipped with multiple antennas ($T_1=4$ and $T_2=2$) and all users are equipped with a single antenna, i.e., the MISO case.}
	\label{Fig:CovProbMISO}
\end{figure*}

In this subsection, some simulation results regarding to the coverage probability are presented. Our objective here is to numerically verify the coverage performances for the COA and ROA schemes in a two-tier mmWave HetNet where the first tier consists of the macrocell BSs and the second tier consists of the mmWave picocell BSs of 73 GHz. For the COA scheme, the user association for BS $X_{m,i}$ is $\Psi_{m,i}(\|X_{m,i}\|)=\frac{P_mG_{m,i}}{\|X_{m,i}\|^{\alpha_{m,i}}}$ whereas $\Psi_{m,i}(\|X_{m,i}\|)=(\frac{P_mG_{m,i}}{\|X_{m,i}\|^{\alpha_{m,i}}})^{\omega_{m}}$ with $\omega_{m}=W_{\mu}\mathds{1}(m\neq M)+W_{\varepsilon}\mathds{1}(m= M)$ is used for the ROA scheme. All network parameters for simulation are listed in Table \ref{Tab:SimPara}. Note that all the following simulation results are obtained by assuming that the penetration loss of the  mmWave picocells is so large that the NLOS signals from any mmWave picocells are too weak to be detected or received by the users.  

In Fig. \ref{Fig:CovProbSISO}, we present the simulation results of the coverage probabilities for  the COA and ROA schemes and assuming all BSs only have a single antenna, i.e., the single-input-single-output (SISO) case is considered. Here we use a simple unified channel model for the LOS and NLOS channels in the UHF band and this model is based on the 3GPP path loss channel model with $\overline{\alpha}=\widetilde{\alpha}=\alpha_{\mu}=3.76$ and $\overline{\rho}^2_1=\widetilde{\rho}^2_1=\rho^2_1=13$ dB\cite{36.872}. In Fig. \ref{Fig:CovProbSISO}, the analytical result of $p_{cov}(\theta)$ based on this unified channel model for the COA scheme is calculated by \eqref{Eqn:MaxCovProb} whereas the analytical result of $p_{cov(\theta)}$ for the ROA scheme is calculated based on the result in Corollary \ref{Cor:CoverageROA}. As can be seen, the analytical $p_{cov}(\theta)$ and the simulated $p_{cov}(\theta)$ are fairly close to each other (their difference is below 2\% on average), which validates that the approximated expression of $p_{cov}(\theta)$ in Theorem \ref{Thm:CoverageProbBias} is very accurate.  The coverage probability $p_{cov}(\theta)$ decreases as the ratio of the mmWave picocell intensity to the blockage intensity increases. This is because the interference in the mmWave band increases as the picocell BSs are deployed faster than the blockages so that users get closer to the interfering picocell BSs and they thus are more likely to get LOS interference channels from the picocell BSs. In addition, how the intensity of the macrocell BSs affects $p_{cov}(\theta)$ can also be observed in Fig. \ref{Fig:CovProbSISO} even though we merely use the fixed intensity of the macrocell BSs to obtain the simulation results. In Fig. \ref{Fig:CovProbSISO}, the curve for 1-tier mmWave BSs is the case when the intensity of the macrocell BSs is zero, i.e., the network only has one-tier mmWave BSs. In other words, if the intensity of the macrocell BSs decreases, the coverage probability will decrease as well since the users inside/behind the blockages that cannot be covered by the mmWave picocell BSs have less chance to be covered by the macrocell BSs. The same reasoning can be applied to understand why the coverage probability improves when the intensity of the macrocell BSs increases. As such, $p_{cov}(\theta)$ of 1-tier mmWave BSs is much smaller than that of the 2-tier mmWave HetNet because the users in the blockages can still be covered by the macrocell BSs. As $\frac{\lambda_2}{\beta}$ goes to infinity, we can expect that $p_{cov}(\theta)$ of the 1-tier mmWave picocell network will converge to around 0.6, which means there are about 40\% of users that are not well covered by the stand-alone mmWave picocell BSs and this portion of users eventually needs to be covered by the UHF macrocell BSs.  From Fig. \ref{Fig:CovProbSISO}, we thus can know that jointly deploying UHF and mmWave BSs in a network indeed significantly improves the network coverage. Also, we can see that the COA scheme achieves a higher coverage probability than the ROA scheme, which coincides our previous discussion in Section \ref{SubSec:CovProbRateCOA}, that is, the COA scheme always outperforms the ROA scheme in terms of the coverage probability.

For the MISO case, the simulation results of $p_{cov}(\theta)$ for the COA and ROA schemes are shown in Fig. \ref{Fig:CovProbMISO} by considering all macrocell BSs equipped with 4 transmit antennas and picocell BSs equipped with 2 transmit antennas. In the figure, all coverage probabilities are significantly improved due to multiple transmit antennas if compared with their corresponding results in Fig. \ref{Fig:CovProbSISO}. The phenomena shown in Fig. \ref{Fig:CovProbMISO} are very similar to those shown in Fig.  \ref{Fig:CovProbSISO} and the analytical results of $p_{cov}(\theta)$ are also very close to their corresponding simulated results. As can be seen in Fig. \ref{Fig:CovProbMISO}, the COA scheme still outperforms the ROA scheme in terms of coverage, but not very much. The coverage gap between COA and ROA is largely reduced due to multiple transmit antennas. As a result, user association may not play a pivotal role in affecting the coverage any more if BSs have multiple antennas. Furthermore, the MISO simulation results in Fig. \ref{Fig:CovProbMISO} are obtained by the assumption that perfect channel state information (CSI) of users is available at each BS, whereas the SISO simulation results in  Fig. \ref{Fig:CovProbMISO} are equivalent to the simulation results in the scenario where no CSI is available at each BS with multiple antennas in that transmitters cannot exploit the transmit diversity if they do not have CSI. In other words, comparing Fig. \ref{Fig:CovProbMISO} with Fig. \ref{Fig:CovProbSISO} is able to help us get some insights into how much CSI available at BSs impacts the performance of the coverage probability. 

\subsection{Numerical Results of the Link Rate}\label{SubSec:NumResultErgodicRate}
\begin{figure*}[!t]
	\centering
	\includegraphics[width=\textwidth]{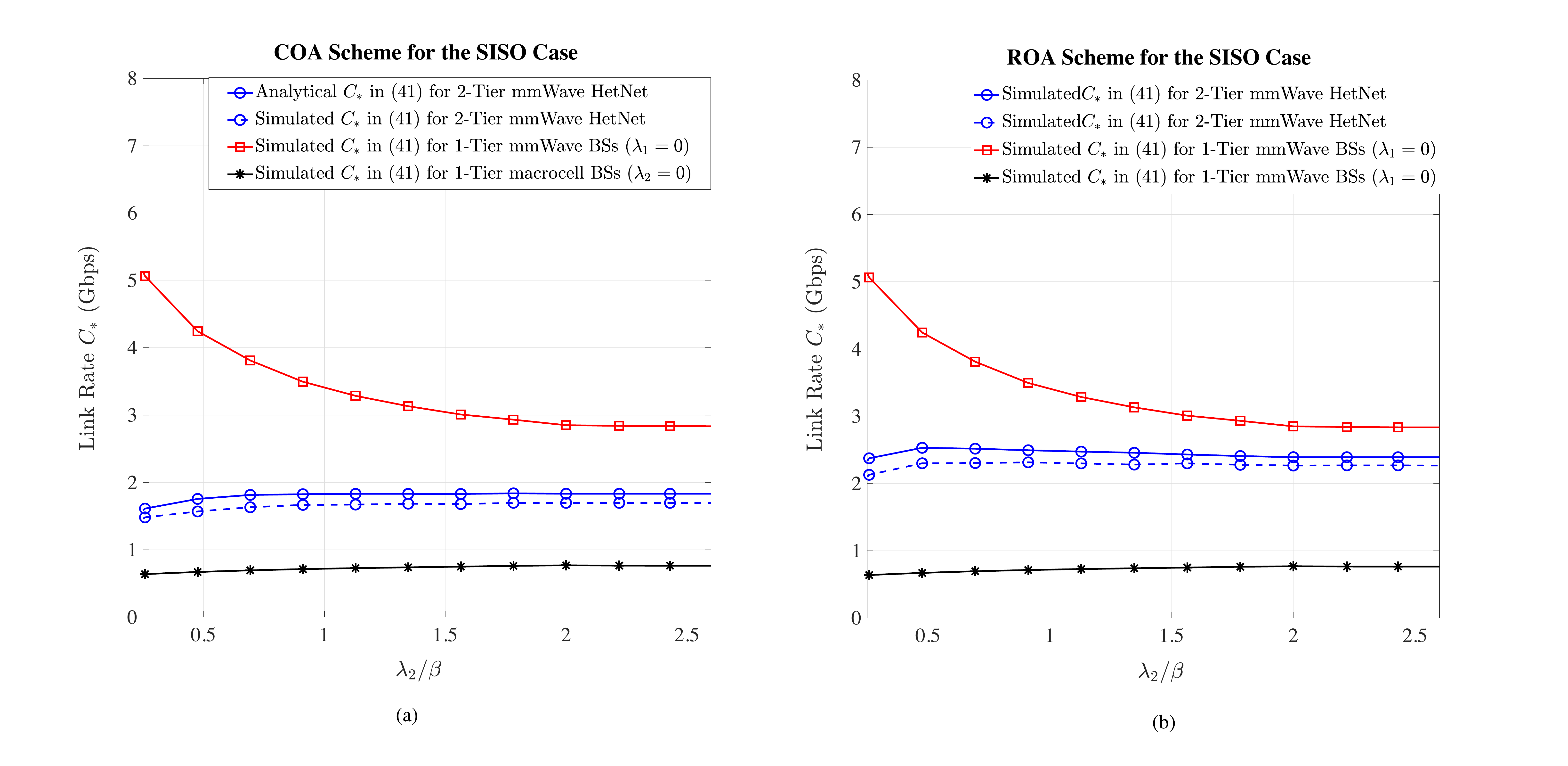}
	\caption{Simulation results of the link rates (a) the COA scheme (b) the ROA scheme. In this simulation, all BSs and users are equipped with a single antenna, i.e., the SISO case.}
	\label{Fig:LinkRateSISO}
\end{figure*}

\begin{figure*}[!t]
	\centering
	\includegraphics[width=\textwidth]{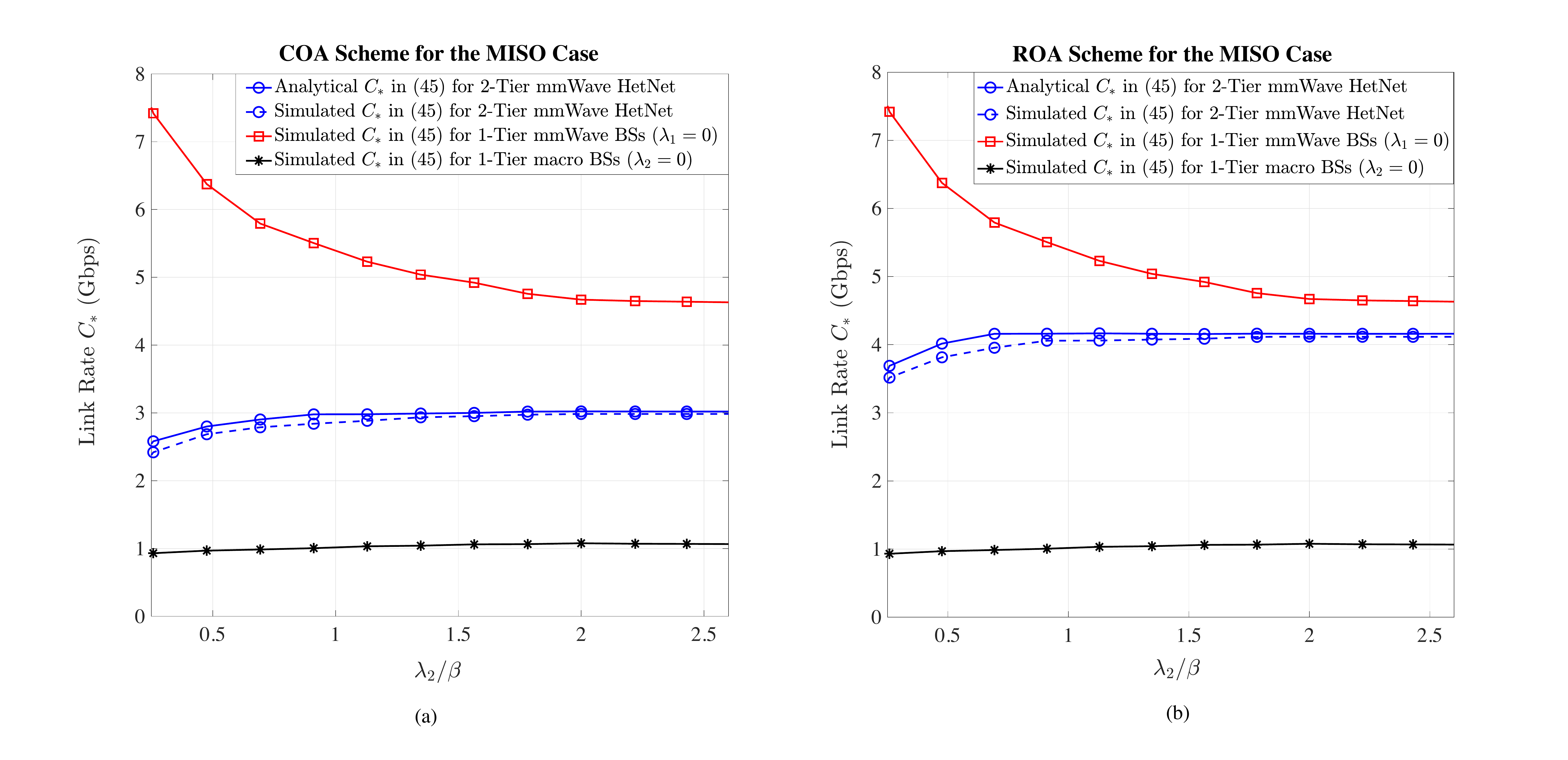}
	\caption{Simulation results of the link rates (a) the COA scheme (b) the ROA scheme.  In this simulation, all BSs are equipped with multiple antennas ($T_1=4$ and $T_2=2$) and users are equipped with a single antenna, i.e., the MISO case.}
	\label{Fig:LinkRateMISO}
\end{figure*}

In this subsection, we show some numerical results of the link rates for the COA and ROA schemes and our goal here is to visually demonstrate how much of the link rates can be achieved by the COA and ROA schemes. All network parameters and assumptions for the simulations here are the same as those in Table \ref{Tab:SimPara}. The link rate of the SISO case is shown in Fig. \ref{Fig:LinkRateSISO}, whereas  the link rate of the MISO case is shown in Fig. \ref{Fig:LinkRateMISO}.

First of all, let us discuss what we can observe and learn from the two (blue) curves of the 2-tier mmWave HetNet in Fig. \ref{Fig:LinkRateSISO}. We can observe that the two (blue)  curves of the mmWave HetNet case initially increase and then slightly decrease and eventually converge to a constant as more and more mmWave BSs are deployed. Their initial increase is thanks to the increase in the SINR by deploying more mmWave BSs, however, deploying too many mmWave BSs eventually results in their decrease due to too much interference. Also, we can see that there exists an optimal value of $\frac{\lambda_2}{\beta}$ that maximizes $C_*$. In the ROA case of 2-tier mmWave HetNet, for instance, $C_*$ maximizes when $\frac{\lambda_2}{\beta}\approx 0.45$. Moreover, there are two more interesting phenomena that can be observed from these two curves. (i)  the approximated analytical results of $C_*$ which are found based on the result in \eqref{Eqn:LinkRateCOA} and the result in \eqref{Eqn:LinkRateROA} by using the unified path loss model for LOS and NLOS channels in the UHF band  for $\overline{\alpha}=\widetilde{\alpha}=\alpha_{\mu}=3.76$ and $\overline{\rho}^2_1=\widetilde{\rho}^2_1=\rho^2_1=13$ dB, are pretty accurate since they are very close to their corresponding simulated results; (ii) ROA significantly outperforms COA in terms of link rate (ROA can improve the link rate of users rate by 30\% on average if compared with COA). Hence, comparing Fig. \ref{Fig:CovProbSISO} and Fig. \ref{Fig:LinkRateSISO} validates the coverage-rate tradeoff problem that indeed exists in a mmWave HetNet\footnote{ Also, please note that the coverage-rate tradeoff problem indeed exists in a mmWave HetNet with different available spectra no matter which values of the network parameters are used for simulation, which can be concluded from Lemmas \ref{Lem:COA} and \ref{Lem:ROA}.}.

There are two other curves in Fig. \ref{Fig:LinkRateSISO} that need to be addressed. In Fig. \ref{Fig:LinkRateSISO}, the (red) curve for  mmWave picocell BSs is the link rate for the network only consisting of mmWave picocell BSs, i.e., no macrocell BSs in the network or $\lambda_1=0$. We can see that this (red) curve is much higher than other three curves because in this one-tier mmWave network users (except the users in the blockages that cannot connect to mmWave BSs) connect to mmWave BSs with a large mmWave bandwidth so that the link rate of the users on average is very high even though some users in the blockages have a zero link rate. The (black) curve represents another extreme case of the network only consisting of the macrocell BSs, i.e., no mmWave picocell BSs in the network or $\lambda_2=0$. This (black) curve is much lower than the other three curves since in this network all users connect to macrocell BSs with a narrow UHF bandwidth. Hence, when mmWave picocell BSs start to be deployed in the network, the (black) curve will move up to the position of the blue curves, whereas it will further move up to the position of the (red) curve as the intensity of the macrocell BSs reduces to zero. Please note that although the 1-tier mmWave network can achieve the highest link rate among the four kinds of networks its coverage probability performance, as shown Fig. \ref{Fig:CovProbSISO}, is the worst among the four. The simulation results of the link rate for the MISO case are shown in Fig. \ref{Fig:LinkRateMISO}. They are much better than those in Fig. \ref{Fig:CovProbSISO} owing to the transmit diversity exploited by each BS with perfect CSI, and they also reveal some implications similar to those observed in Fig. \ref{Fig:LinkRateSISO}. Without perfect CSI on the transmitter side, the results in Fig. \ref{Fig:LinkRateMISO} downgrade to their corresponding results in Fig. \ref{Fig:LinkRateSISO}. If we compare Fig. \ref{Fig:LinkRateSISO}(b) with Fig. \ref{Fig:LinkRateMISO}(b), we can find a subtle difference between the (blue) curves in these two figures, that is, the link rate in the SISO case does not always increase as $\lambda_2/\beta$ increases, which is different from the link rate in the MSIO case. This subtle difference is coming from the fact that SISO channels are much more sensitive to the blockage environment than MISO channels. The link rate of MISO channels is mainly impacted by whether the transmit diversity can be exploited in channels rather than whether the channels are blocked. Finally, we can summarize that the simulation results in Figs. \ref{Fig:CovProbSISO}-\ref{Fig:LinkRateMISO} not only verify that there indeed exists the coverage-rate tradeoff problem in a mmWave HetNet, but also suggest that the tradeoff problem could be alleviated by the multi-antenna communication techniques that are able to improve the coverage and link rate at the same time.

\section{Conclusion}\label{Sec:Conclusion}
In an urban area, the characteristics of wireless channels are seriously affected by the blockages, especially the channels in the mmWave band. To completely characterize LOS and NLOS channels induced by the blockages, in this work we develop a very general modeling and analysis approach based on stochastic geometry to fundamentally characterize the relationships between user association, coverage probability and link rate. The general expressions of the coverage probability and link rate for the GUA scheme are approximately derived in a compact form that straightforwardly indicates how LOS and NLOS channels, user association parameters, blockage intensity and MISO fading affect the coverage probability as well as the link rate. Most importantly, they shed light on the fundamental tradeoff problem between coverage and link rate that exists in a mmWave HetNet with different bandwidths in the UHF and mmWave spectra. We characterize how to design the user association functions for the COA and ROA schemes and show that simultaneously maximizing the coverage and link rate only can be achieved when there is no bandwidth discrepancy.     

\appendix 

\subsection{Proof of Theorem \ref{Thm:CDFAssBS}}\label{App:ProofCDFAssBS}
The CDF of $\Psi_*(\|X_*\|)$, $F_{\Psi_*(\|X_*\|)}(x)$, can be written as
\begin{align*}
 &\mathbb{P}\left[\Psi_*(\|X_*\|)\leq x\right]=\mathbb{P}\left[\sup_{m,i:X_{m,i}\in\Phi} \Psi_{m,i}(\|X_{m,i}\|)\leq x\right]\\
\stackrel{(a)}{=}&\prod_{m=1}^{M}\mathbb{E}_{\Phi_m}\left\{\prod_{m,i\in\Phi_m}\mathbb{P}\left[\Psi_{m,i}(\|X_{m,i}\|)\leq x| \Phi_m\right]\right\}\\
 \stackrel{(b)}{=}& \exp\left(-2\pi\sum_{m=1}^{M}\lambda_m\int_{\mathbb{R}_+}\mathbb{P}[\Psi_m(r)>x]r\dif r\right)\\
=&\exp\bigg(-2\pi\sum_{m=1}^{M}\lambda_m\int_{\mathbb{R}_+}\mathbb{P}[\ell(r)\overline{\Psi}_{m}(r)+(1-\ell(r))\\
&\times\widetilde{\Psi}_{m}(r)>x]r\dif r\bigg)\\
\stackrel{(c)}{=}&\exp\bigg(-2\pi\sum_{m=1}^{M}\lambda_m\int_{\mathbb{R}_+}\big\{e^{-\eta\beta r}\big(\mathbb{P}[\overline{\Psi}_{m}(r)>x]\\
&-\mathbb{P}[\widetilde{\Psi}_m(r)>x]\big)+\mathbb{P}[\widetilde{\Psi}_m(r)>x]\big\}r\dif r\bigg)\\
\stackrel{(d)}{=}&\exp\bigg(-\pi\sum_{m=1}^{M}\lambda_m\int_{\mathbb{R}_+}\bigg\{e^{-\eta\beta r}\bigg(\mathbb{P}\left[r<\overline{\Psi}^{-1}_{m}(x)\right]\\
&-\mathbb{P}\left[r<\widetilde{\Psi}^{-1}_m(x)\right]\bigg)+\mathbb{P}\left[r<\widetilde{\Psi}^{-1}_m(x)\right]\bigg\}\dif r^2\bigg).
\end{align*}
where $(a)$ follows from the independence among all $\Psi_{m,i}(\cdot)$'s, $(b)$ is due to the probability generation functional (PGL) of $M$ independent homogeneous PPPs, $(c)$ follows from $\mathbb{P}[\ell(r)=1]=\exp(-\eta\beta r)$, and $(d)$ follows the assumption that $\overline{\Psi}_{m,i}(\cdot)$ and $\widetilde{\Psi}_{m,i}(\cdot)$ are a bijective (invertible) and monotonic decreasing function. Since we know that for any constant $a>0$ we have the following
\begin{align*}
\int_{0}^{\infty}e^{-ar}\mathbb{P}[r<Z]\dif r^2&=2\mathbb{E}_Z\left[\int_{0}^{\infty}e^{-a r}\mathbb{P}[r<Z|Z]r\dif r\right]\\
&=2\mathbb{E}_Z\left[\int_{0}^{Z}e^{-a r}r\dif r\right],
\end{align*} 
the result in $(d)$ can be expressed as \eqref{Eqn:CDFAssGainBS} with $\mathsf{A}_m(x)$ given in \eqref{Eqn:ExpressionAkm1}. 

The probability that $X_*$ belongs to tier $m$ can be derived as follows. First, we notice that $\mathsf{A}_m(z)$ can be alternatively expressed as 
\begin{align*}
\mathsf{A}_m(z) &=\mathbb{E}\bigg\{\int_{0}^{\infty}2\bigg(\ell(r)\mathbb{P}\left[r<\overline{\Psi}^{-1}_m(z)\right]\\
&+(1-\ell(r))\mathbb{P}\left[r<\widetilde{\Psi}^{-1}_m(z)\right]\bigg)r\dif r \bigg\}.
\end{align*}
Thus, $\mathsf{A}_m\circ\Psi_m(x)\defn \mathsf{A}_m(\Psi_m(x))$ and $\mathsf{A}_k\circ\Psi^{\dagger}_m(x)\defn \mathsf{A}_k(\Psi^{\dagger}_m(x))$ can be found as shown in the following:
\begin{align*}
\mathsf{A}_m\circ\Psi_m(x)=&\mathbb{E}\bigg\{\int_{0}^{\infty}2\bigg(\ell(r)\mathbb{P}\left[r<\overline{\Psi}^{-1}_m\circ\Psi_m(x)\right]\\
&+(1-\ell(r))\mathbb{P}\left[r<\widetilde{\Psi}^{-1}_m\circ\Psi_m(x)\right]\bigg)r\dif r \bigg\}\\
=&\int_{0}^{\infty} 2\mathbb{P}[r<x]r\dif r=x^2,
\end{align*}
\begin{align*}
\mathsf{A}_k\circ\Psi^{\dagger}_m(x)=&\mathbb{E}\bigg\{\int_{0}^{\infty}2\bigg(\ell(r)\mathbb{P}\left[r<\overline{\Psi}^{-1}_k\circ\overline{\Psi}^{\dagger}_m(x)\right]\\
&+(1-\ell(r))\mathbb{P}\left[r<\widetilde{\Psi}^{-1}_k\circ\widetilde{\Psi}^{\dagger}_m(x)\right]\bigg)r\dif r \bigg\},
\end{align*}
which can be shown to equal to the result in \eqref{Eqn:ExpressionAkm3}.  Next, we know that probability $\phi_m$ can be explicitly defined as
\begin{align*}
\phi_m\defn &\mathbb{P}\bigg[\sup_{m,i:X_{m,i}\in\Phi_m} \Psi_{m,i}(\|X_{m,i}\|)>\\
 &\sup_{k,i:X_{k,i}\in\Phi\setminus\Phi_m}\Psi_{k,i}(\|X_{k,i}\|)\bigg]\\
=&\int_{0}^{\infty} \mathbb{P}\left[\sup_{k,i:X_{k,i}\in\Phi\setminus\Phi_m}\Psi_{k,i}(\|X_{k,i}\|)<z\right]\dif F_{Z_m}(z),
\end{align*}
where RV $Z_m\defn \sup_{m,i:X_{m,i}\in\Phi_m} \Psi_{m,i}(\|X_{m,i}\|)$. The CDF of $Z_m$ given by
\begin{align}\label{Eqn:CDFAssFunTierM}
F_{Z_m}(z)=\exp\bigg(-\pi\lambda_m \mathsf{A}_m(z)\bigg),
\end{align}
which can be inferred from \eqref{Eqn:CDFAssGainBS} with only one PPP.  According to \eqref{Eqn:CDFAssGainBS}, we also know
\begin{align}\label{Eqn:CDFAssFunNoTierM}
&\mathbb{P}\left[\sup_{k,i:X_{k,i}\in\Phi\setminus\Phi_m}\Psi_{k,i}(\|X_{k,i}\|)<z\right]\nonumber\\
&=\exp\left(-\pi\sum_{k\in\mathcal{M}\setminus m}\lambda_k \mathsf{A}_k(z)\right).
\end{align}
Thus, substituting \eqref{Eqn:CDFAssFunTierM} and \eqref{Eqn:CDFAssFunNoTierM} into $\phi_m$ given above yields
\begin{align*}
\phi_m=&\bigintsss_{0}^{\infty} e^{-\pi\sum_{k\in\mathcal{M}\setminus m}\lambda_k \mathsf{A}_k(z)}\dif F_{Z_m}(z)\\
=&\mathbb{E}\bigg\{\bigintsss_{0}^{\infty} \exp\left(-\pi\sum_{k\in\mathcal{M}\setminus m}\lambda_k \mathsf{A}_k(\Psi^{\dagger}_m(x))\right)\times\\
& \dif F_m(\Psi^{\dagger}_m(x)) \bigg\}\\
\stackrel{(e)}{=}& 2\pi\lambda_m\mathbb{E}\left\{\bigintsss_{0}^{\infty} \exp\left(-\pi\sum_{k=1}^M\lambda_k \mathsf{A}_k\circ\Psi^{\dagger}_m(x)\right)x \dif x\right\}
%&=2\pi\lambda_m\mathbb{E}\left\{\bigintsss_{0}^{\infty} \exp\left(-\pi\sum_{k=1}^M\lambda_k \mathsf{A}_{km^{\dagger}}(x)\right) x\dif x \right\},
\end{align*}
where $(e)$ follows from $\dif F_m(\Psi^{\dagger}_m(x))=2\pi x\exp(-\pi\lambda_m \mathsf{A}_m\circ\Psi^{\dagger}_m(x))\dif x$ for given $\Psi^{\dagger}_m(x)$. Hence, $\phi_m$ in \eqref{Eqn:GenAssProbTierM} is obtained.

\subsection{Proof of Theorem \ref{Thm:CoverageProbBias}}\label{App:ProofCoverageProbBias}
Since $\Psi_{m,i}(x)=\frac{\omega_mG_{m,i}}{\nu_mx^{\alpha_{m,i}}}$, we know $\overline{\psi}_{m,i}= \frac{\omega_m\overline{G}_{m,i}}{\nu_m}$ and $\widetilde{\psi}_{m,i} = \frac{\omega_m\widetilde{G}_{m,i}}{\nu_m}$.  According to \eqref{Eqn:PowerLawExpressionAkm2}, we can have $\mathsf{A}_k\circ\Psi^{\dagger}_m(x)$ as given in \eqref{Eqn:AkPsimBias} due to $\mathbb{E}[\widetilde{G}^{2/\widetilde{\alpha}}_k]=e^{4\frac{\widetilde{\rho}^2_k}{\widetilde{\alpha}^2}}$. Then substituting this into \eqref{Eqn:GenAssProbTierM} leads to $\phi_m$ in \eqref{Eqn:AssProbTiermBias}.  For $X_*\in \Phi_m$ and $m\in\{1,\ldots, M-1\}$, the coverage probability can be found by
\begin{align*}
&\mathbb{P}\left[\frac{P_*H_*}{I_{*,\mu}L_*(\|X_*\|)}\geq\theta\right]=\mathbb{P}\left[h_{*,\mu} \geq \theta \frac{ I_{*,\mu}L_*(\|X_*\|)}{P_*G_*}\right]\\
&\stackrel{(a)}{=}\sum_{n=0}^{T_m-1}\frac{(-\theta)^n}{n!}\frac{\dif^n}{\dif \theta^n}\mathbb{E}\left[e^{- \frac{ \theta \omega_*I_{*,\mu}L_*(\|X_*\|)}{P_*\omega_*G_*}}\right],
\end{align*}
where $(a)$ follows from $\mathbb{P}[Z\geq \theta z]=\sum^{T_m-1}_{n=0}\frac{(\theta z)^n}{n!}e^{-\theta z}=\sum^{T_m-1}_{n=0}\frac{(-\theta)^n}{n!}\frac{\dif^n}{\dif\theta^n}e^{-\theta z}$ if $Z$ is a Chi-square RV with $2T_m$ degrees of freedom, $\frac{\omega_*G_*}{L_*(\|X_*\|)}\defn \sup_{m,i:X_{m,i}\in\Phi} \frac{\omega_{m}G_{m,i}}{L_{m,i}(\|X_{m,i}\|)}$ and $\omega_*\in\{\omega_m, m\in\mathcal{M}\}$ is the user association bias used by BS $X_*$. 

According to \eqref{Eqn:PowerLawUserAssFun},  the probability of $\frac{\omega_*G_*}{L_*(\|X_*\|)}\leq \frac{1}{L_*(x)}$ can be written as
\begin{align*}
&\mathbb{P}\left[\frac{\omega_*G_*}{L_*(\|X_*\|)}\leq\frac{1}{L_*(x)}\right]=\mathbb{P}\left[\frac{\|X_*\|}{\left(\omega_*G_*/\nu_*\right)^{\frac{1}{\alpha_*}}}\geq x\right]\\
&=\exp\left\{-\pi\sum_{m=1}^{M}\lambda_m\mathsf{A}_m\left(x\right)\right\},
\end{align*}
where $\nu_*\in\{\nu_{\mu},\nu_{\varepsilon}\}$ is the intercept used by $X_*$ and $\mathsf{A}_m\left(x\right)$ is found by
\begin{align*}
\mathsf{A}_m(x)=& x^2  \bigg(\mathbb{E}\left[\int_{(\omega_m\widetilde{G}_m/\nu_m)^{1/\widetilde{\alpha}}}^{(\omega_m\overline{G}_m/\nu_m)^{1/\overline{\alpha}}} 2te^{-\eta\beta x t}\dif t\right]\\
&+\left(\frac{\omega_m}{\nu_m}\right)^{\frac{2}{\widetilde{\alpha}}}\mathbb{E}\left[\widetilde{G}_m^{\frac{2}{\widetilde{\alpha}}}\right]\bigg)\\
=& 2\int_{0}^{x} \mathbb{E}\bigg[\left(\frac{\omega_m\overline{G}_m}{\nu_m}\right)^{2/\overline{\alpha}}e^{-\eta\beta(\omega_m\overline{G}_m/\nu_m)^{1/\overline{\alpha}} r}\\
&+\left(\frac{\omega_m\widetilde{G}_m}{\nu_m}\right)^{2/\widetilde{\alpha}}\left(1-e^{-\eta\beta(\omega_m\widetilde{G}_m/\nu_m)^{1/\widetilde{\alpha}} r}\right)\bigg]  r\dif r
\end{align*}
because we let $\Psi_{m,i}(x)=x(\omega_mG_{m,i}/\nu_m)^{-1/\alpha_{m,i}}$ and $\Psi^{-1}_{m,i}(x)=x(\omega_mG_{m,i}/\nu_m)^{1/\alpha_{m,i}}$ and then substitute them into \eqref{Eqn:ExpressionAkm1}. Thus, it follows that
\begin{align*}
\mathbb{P}\left[\frac{\|X_*\|}{\left(\frac{\omega_*G_*}{\nu_*}\right)^{\frac{1}{\alpha_*}}}\geq x\right]=\exp\left\{-2\pi\int_{0}^{x}\sum_{k=1}^{M}\Lambda_k\left(0,0,r\right)r\dif r\right\},
\end{align*}
where $\Lambda_k(\cdot,\cdot,\cdot) $ is given in \eqref{Eqn:TiermIntensityBias} and this manifests that $X_*(\omega_*G_*/\nu_*)^{-\frac{1}{\alpha_*}}$ can be viewed as the point of an inhomogeneous PPP with intensity $\Lambda(0,0,r)=\sum_{k=1}^{M}\Lambda_k(0,0,r)$ nearest to the typical user. Besides, we have
\begin{align*}
&\mathbb{E}\left[\exp\left(- \frac{ \theta \omega_*I_{*,\mu}L_*(\|X_*\|)}{P_*\omega_*G_*}\right)\right]\\
&\stackrel{(b)}{=}\mathbb{E}\left[\exp\left(-\frac{ \theta \omega_*}{P_*}\sum_{k,i:\widehat{X}_{k,i}\in\widehat{\Phi}\setminus\widehat{X}_*} \frac{P_kH_{k,i}\|\widehat{X}_*\|^{\alpha_*}}{\omega_k\|\widehat{X}_{k,i}\|^{\alpha_k }}\right)\right]\\
&\stackrel{(c)}{\approx}\mathbb{E}\left[\exp\left(-\frac{ \theta \omega_*}{P_*}\sum_{k,i:\widehat{X}_{k,i}\in\widehat{\Phi}\setminus\widehat{X}_*} \frac{P_kH_{k,i}\|\widehat{X}_*\|^{\alpha_*}}{\omega_k\|\widehat{X}_{k,i}\|^{\alpha_*}}\right)\right],
\end{align*}
where $(b)$ follows from the result of Theorem 1 in \cite{CHLLCW16} by letting $\widehat{\Phi}\defn \bigcup_{m=1}^{M-1} \widehat{\Phi}_m$, $\widehat{\Phi}_m$ is an inhomogeneous PPP of intensity $\Lambda_m(0,0,r)$ and $\widehat{X}_*\in\widehat{\Phi}$ is the nearest BS to the typical user. The approximation in $(c)$ is made by letting the path loss exponents of all interfering $\widehat{X}_{m,i}$'s be equal to the exponent of $\widehat{X}_*$ in order to facilitate the following derivations\footnote{Since $\widehat{X}_*$ is the nearest BS to the typical user, a few interfering BSs close to $\widehat{X}_*$ would be in a blockage environment similar to $\widehat{X}_*$ so that they would have a path loss exponent close to the path loss exponent of $\widehat{X}_*$. For the BSs far away from $\widehat{X}_*$, they can also use the path loss exponent of  $\widehat{X}_*$ as their path loss exponent since in general they do not contribute too much interference to the typical user and changing their path loss exponents does not affect the interference too much. Thus, the approximation in $(c)$ is usually accurate and it becomes exact as $\beta=0$ and $\beta=\infty$. }. Therefore, we can have
\begin{align*}
&\mathbb{E}\left[\exp\left(-\frac{ \theta \omega_*}{P_*}\sum_{k,i:\widehat{X}_{k,i}\in\widehat{\Phi}\setminus\widehat{X}_*} \frac{P_kH_{k,i}\|\widehat{X}_*\|^{\alpha_*}}{\omega_k\|\widehat{X}_{k,i}\|^{\alpha_*}}\right)\right]\\
&=\mathbb{E}\left[\exp\left(-\frac{ \theta \omega_*}{P_*}\sum_{k,i:\widehat{X}_{k,i}\in\widehat{\Phi}} \frac{P_kH_{k,i}}{\omega_k}\left(\frac{\|\widehat{X}_{k,i}\|^{2}}{\|\widehat{X}_*\|^{2}}\right)^{-\frac{\alpha_*}{2}}\right) \right]\\
&=\prod_{k=1}^{M-1}\mathbb{E}_{\widehat{\Phi}_k}\left\{\prod_{k,i:\widehat{X}_{k,i}\in\widehat{\Phi}_k}\mathbb{E}\left[e^{-\frac{ \theta \omega_*P_k}{P_*\omega_k} H_{k,i}\left(\frac{\|\widehat{X}_{k,i}\|^{2}}{\|\widehat{X}_*\|^{2}}\right)^{-\frac{\alpha_*}{2}}}\right]\right\}\\
&=\prod_{k=1}^{M-1}\mathbb{E}_{\widehat{\Phi}_k}\left\{\prod_{\widehat{X}_{k,i}\in\widehat{\Phi}_k} \frac{1}{1+\frac{ \theta \omega_*P_k}{P_*\omega_k}\left(\frac{\|\widehat{X}_*\|^{2}}{\|\widehat{X}_{k,i}\|^{2}}\right)^{\frac{\alpha_*}{2}}}\right\}\\
&\stackrel{(d)}{\approx}\exp\left(-2\pi \sum_{k=1}^{M-1}\int^{\infty}_x\Lambda_k\left(\frac{r}{x},\frac{P_*}{\theta\omega_*},r\right)r\dif r \right)\\
%&\stackrel{(b)}{=}\exp\left(-\pi x^2\sum_{k=1}^{M-1}\int^{\infty}_1 \Lambda_k(\sqrt{u}x)\left[\frac{e^{-\eta\beta \sqrt{u}x}}{u^{\overline{\alpha}/2}\frac{P_*\omega_k}{ \theta \omega_*P_k}+1}+\frac{\left(1-e^{-\eta\beta \sqrt{u}x}\right)}{u^{\widetilde{\alpha}/2}\frac{P_*\omega_k}{ \theta \omega_*P_k}+1}\right]\dif u \right),
&=\exp\left(-\pi x^2 \sum_{k=1}^{M-1} \int^{\infty}_1\frac{\Lambda_k(\sqrt{u}x)}{u^{\frac{\alpha}{2}}\frac{P_*\omega_k}{ \theta b_*P_k}+1}\dif u \right),
\end{align*}
where $(d)$ follows by letting $\|\widehat{X}_*\|=x$ and assuming $\widehat{\Phi}$ consists of $m$ independent inhomogeneous PPPs and then using the probability generating functional (PGFL) of $\widehat{\Phi}$ \cite{DSWKJM13} for given $\|\widehat{X}_*\|=x$,  $\mathbb{P}[\alpha_*=\overline{\alpha}]=\exp(-\eta\beta r)$  as well as $\mathbb{P}[\alpha_*=\widetilde{\alpha}]=1-\exp(-\eta\beta r)$.
If $X_*\in\Phi_m$ and $f_{\|\widehat{X}_*\|}(x)=2\pi x\Lambda(x)e^{-2\pi \int_{0}^{x}\Lambda(0,0,r)r\dif r}$, we can have 
\begin{align*}
&\mathbb{E}\left[\exp\left(- \frac{ \theta I_{*,\mu}L_*(\|X_*\|)}{P_*H_*}\right)\bigg| X_*\in\Phi_m\right]\\
\approx& \int_{0}^{\infty}\exp\left(-\pi x^2 \sum_{k=1}^{M-1} \int^{\infty}_1\frac{\Lambda_k(\sqrt{u}x)}{u^{\frac{\alpha}{2}}\frac{P_*\omega_k}{ \theta b_*P_k}+1}\dif u \right)\times\\
 &f_{\|\widehat{X}_*\|}(x)\dif x
\defn \mathsf{B}_m(\theta).
\end{align*}
Thus, carrying out the above integral yields $\mathsf{B}_m(\theta)$ in \eqref{Eqn:BmfunBias}.
Similarly, if $X_*\in\Phi_M$, the coverage probability can be shown as
\begin{align*}
&\mathbb{P}\left[\frac{P_*H_*/L_*(\|X_*\|)}{I_{*,\varepsilon}+\sigma^2_{\varepsilon}}\geq \theta\right]\\
&=\mathbb{P}\left[h_{*,\varepsilon}\geq \frac{\theta(I_{*,\varepsilon}+\sigma^2_{\varepsilon})}{T_MP_MG_*/L_*(\|X_*\|)}\right]\\
&=\mathbb{E}\left[\exp\left(-\frac{\theta(I_{*,\varepsilon}+\sigma^2_{\varepsilon})}{T_MP_MG_*/L_*(\|X_*\|)}\right)\right],
\end{align*}
and using the same approach of showing $\mathsf{B}_m$ can help us show that
 $\mathbb{E}\left[\exp\left(-\frac{\theta(I_{*,\varepsilon}+\sigma^2_{\varepsilon})L_*(\|X_*\|)}{T_MP_MG_*}\right)\right]\approx \mathsf{B}_M(\theta),$
which completes the proof. 

\subsection{Proof of Theorem \ref{Thm:ErgodicLinkRate}}\label{App:ProofErgodicLinkRate}
The link rate of the UHF BSs shown in \eqref{Eqn:ErgodicLinkThroughput} can be explicitly written by using the result in Theorem 1 \cite{CHLLCW16} as follows
\begin{align}
C_{m}&=W_{\mu}\mathbb{E}\left[\ln\left(1+\frac{P_mH_m}{I_{*,\mu}L_*(\|X_*\|)}\right)\right]\nonumber\\
&=W_{\mu}\mathbb{E}\left[\ln\left(1+\frac{P_mH_m}{\omega_m\widehat{I}^*_{\mu}L_*(\|\widehat{X}_*\|)}\right)\right], \label{Eqn:ThroughputUHFProof}
\end{align}
where $\widehat{I}^*_{\mu}$ is defined as $\widehat{I}^*_{\mu}\defn \sum_{m,i:\widehat{X}_{m,i}\in\widehat{\Phi}\setminus \widehat{X}_*} \frac{P_{m,i}H_{m,i}}{\omega_mL_{m,i}(\|\widehat{X}_{m,i}\|)}$ in which $\widehat{\Phi}\defn\bigcup_{m=1}^{M-1}\widehat{\Phi}_m$ and $\widehat{\Phi}_m$ is an inhomogeneous PPP of intensity $\Lambda_m(0,0,r)$ that is already defined in Theorem \ref{Thm:CoverageProbBias}. Using the integral identity of the Shannon transformation in Theorem 1 in \cite{CHLHCT17}, $C_m$ in \eqref{Eqn:ThroughputUHFProof} can be further expressed as
\begin{align}\label{Eqn:ErgodicThroughputUHFProof}
C_{m} = \int_{0^+}^{\infty} \frac{1}{s}\left[1-\mathcal{L}_{\frac{P_mH_m}{\omega_m}}(s)\right] \mathcal{L}_{\widehat{I}_{*,\mu}L_*(\|\widehat{X}_*\|)}(s) \dif s,
\end{align}
where $\mathcal{L}_Z(s)\defn \mathbb{E}\left[e^{-sZ}\right]$ is the Laplace transform of nonnegative RV $Z$ for any $s>0$. The Laplace transform of $\frac{P_mH_m}{\omega_m}$ can be found by $\mathcal{L}_{\frac{P_mH_m}{\omega_m}}(s)=  \left(1+\frac{sP_m}{\omega_m}\right)^{-T_m}$ and the Laplace transform of $\widehat{I}^*_{\mu}L_*(\|\widehat{X}_*\|)$  for $\widehat{X}_*\in\widehat{\Phi}_m$  can be found as shown in the following:
\begin{align*}
&\mathcal{L}_{\widehat{I}^*_{\mu}L_*(\|\widehat{X}_*\|)}(s)\\
&= \mathbb{E}\left[\exp\left(-\sum_{k,i:\widehat{X}_{k,i}\in\widehat{\Phi}\setminus \widehat{X}_*} \frac{sP_{k,i}H_{k,i}L_*(\|\widehat{X}_*\|)}{\omega_kL_{k,i}(\|\widehat{X}_{k,i}\|)}\right)\right]\\
&=\mathbb{E}\left[\exp\left(-\sum_{k,i:\widehat{X}_{k,i}\in\widehat{\Phi}\setminus \widehat{X}_*} \frac{sP_{k,i}H_{k,i}\|\widehat{X}_*\|^{\alpha_*}}{\omega_k\|\widehat{X}_{k,i}\|^{\alpha_{k,i}}}\right)\right]
\end{align*}
\begin{align*}
&\stackrel{(a)}{\approx} \mathbb{E}\left[\exp\left(-\sum_{k,i:\widehat{X}_{k,i}\in\widehat{\Phi}\setminus \widehat{X}_*} \frac{sP_{k,i}H_{k,i}\|\widehat{X}_*\|^{\alpha_*}}{\omega_k\|\widehat{X}_{k,i}\|^{\alpha_*}}\right)\right]\\
&\stackrel{(b)}{\approx}\prod_{k=1}^{M-1}\mathbb{E}_{\widehat{\Phi}_k}\left[\prod_{k,i:\widehat{X}_{k,i}\in\widehat{\Phi}\setminus\widehat{X}_*}\frac{1}{1+\frac{ s P_k}{\omega_k}\left(\frac{\|\widehat{X}_*\|^{2}}{\|\widehat{X}_{k,i}\|^{2}}\right)^{\frac{\alpha_*}{2}}}\right] \\
&\stackrel{(c)}{=}\mathsf{B}_m\left(\frac{P_m}{s\omega_m}\right),
\end{align*}
where $(a)$ follows from the reasoning in the proof of Theorem \ref{Thm:CoverageProbBias} in Appendix \ref{App:ProofCoverageProbBias} that changing all the path loss exponents of all interference BSs to the path loss exponent of the serving BSs can give us a good approximation, $(b)$ is due to the assumption that the $m$ inhomogeneous PPP in $\widehat{\Phi}$ are independent, and $(c)$ follows the result in the proof of Theorem \ref{Thm:CoverageProbBias} in Appendix \ref{App:ProofCoverageProbBias} and can be expressed as $\mathsf{B}_m\left(\frac{P_m}{s\omega_m}\right)$. Similarly, $C_M$ can be expressed as
\begin{align}\label{Eqn:ErgodicThroughputmmWaveProof}
C_M&\defn W_{\mu}\mathbb{E}\left[\ln\left(1+\frac{T_MP_MH_M}{\omega_M\widehat{I}_{*,\varepsilon}L_*(\|\widehat{X}_*\|)}\right)\right]\nonumber\\
&=W_{\mu}\int_{0^+}^{\infty}\left[1-\mathcal{L}_{\frac{P_MH_M}{\omega_M}}(s)\right] \mathcal{L}_{\widehat{I}_{*,\varepsilon}L_*(\|\widehat{X}_*\|)}(s)  \frac{\dif s}{s}.
\end{align}
In addition, we can show $\mathcal{L}_{\frac{T_MP_Mh_M}{\omega_M}}(s) = \left(1+\frac{sT_MP_M}{\omega_M}\right)^{-1}$ and $\mathcal{L}_{\widehat{I}^*_{\varepsilon}L_*(\|\widehat{X}_*\|)}(s)=\mathsf{B}_M(1/s)$. Substituting the above results of $\mathcal{L}_{\frac{P_mH_m}{\omega_m}}(s)$ and $\mathcal{L}_{\widehat{I}^*_{\mu}L_*(\|\widehat{X}_*\|)}(s)$ into \eqref{Eqn:ErgodicThroughputUHFProof} and the above results of $\mathcal{L}_{\frac{P_MH_M}{\omega_M}}(s)$ and $\mathcal{L}_{\widehat{I}^*_{\varepsilon}L_*(\|\widehat{X}_*\|)}(s)$ into \eqref{Eqn:ErgodicThroughputmmWaveProof} yields \eqref{Eqn:ErgodicLinkThroughput}.

\subsection{Proof of Lemma \ref{Lem:COA}}\label{App:ProofCOA}
For the COA scheme, BS $X_*$ that provides the maximum coverage to a user can be written as
\begin{align*}
X_*& = \arg \sup_{m,i:X_{m,i}\in\Phi}\mathbb{P}\left[\gamma_{m,i}(\|X_{m,i}\|\geq \theta)\right]\\
&=\arg \sup_{m,i:X_{m,i}\in\Phi}\gamma_{m,i}(\|X_{m,i}\|),
\end{align*}
where $\gamma_{m,i}(\|X_{m,i}\|)$ is the SINR if the typical user associates with BS $X_{m,i}$. Now consider a realization of set $\Phi$ and the total signal power (desired signal power plus interference power plus noise power) received by the typical user is $I_0 \defn I_{0,\mu}\mathds{1}(m\neq M)+I_{0,\varepsilon}\mathds{1}(m=M)$ where $I_{0,\mu}$ and $I_{o,\varepsilon}$ are the total received signal power plus noise power in the UHF and mmWave bands, respectively. Then we can have $\gamma_{m,i}(\|X_{m,i}\|)=\frac{P_mh_{m,i}G_{m,i}\|X_{m,i}\|^{-\alpha_{m,i}}}{I_0-P_mh_{m,i}G_{m,i}\|X_{m,i}\|^{-\alpha_{m,i}}}$ and it follows that
\begin{align*}
X_*&=\arg\sup_{m,i:X_{m,i}\in\Phi} \gamma_{m,i}(\|X_{m,i}\|)\\
&=\arg\inf_{m,i:X_{m,i}\in\Phi} \left(\frac{I_0\|X_{m,i}\|^{\alpha_{m,i}}}{P_mh_{m,i}G_{m,i}}-1\right)\\
&=\arg\sup_{m,i:X_{m,i}\in\Phi} \frac{P_mh_{m,i}G_{m,i}}{\|X_{m,i}\|^{\alpha_{m,i}}}.
\end{align*}
Since $h_{m,i}\sim \exp(1)$ and users only use the mean power of the user association signals to associate with a BS, the COA scheme can be implemented at the user becomes
$X_* = \arg\sup_{m,i:X_{m,i}\in\Phi} \frac{P_m G_{m,i}}{\|X_{m,i}\|^{\alpha_{m,i}}},$
which is exactly the GUA scheme with $\Psi_{m,i}=\frac{\omega_m G_{m,i}}{\|X_{m,i}\|^{\alpha_{m,i}}}$ and $\omega_{m}=P_m$. Hence, using $\Psi_{m,i}=\frac{P_m G_{m,i}}{\|X_{m,i}\|^{\alpha_{m,i}}}$ as a user association function can maximize the coverage probability on average. 

\subsection{Proof of Lemma \ref{Lem:ROA}}\label{App:ProofROA}
For the ROA scheme, if all small-scale fading gains are averaged out at the typical user, BS $X_*$ that provides the typical user with the maximum link rate can be expressed as
\begin{align*}
X_* &= \arg \sup_{m,i:X_{m,i}\in\Phi} \omega_{m}\ln\left[1+\gamma_{m,i}(\|X_{m,i}\|)\right]\\
&= \arg \sup_{m,i:X_{m,i}\in\Phi} \left[\frac{I_0}{I_0-P_mG_{m,i}\|X_{m,i}\|^{-\alpha_{m,i}}}\right]^{\omega_m}\\
&=  \arg \inf_{m,i:X_{m,i}\in\Phi} \left[1-\frac{P_mG_{m,i}}{I_0\|X_{m,i}\|^{\alpha_{m,i}}}\right]^{\omega_m}\\
&\stackrel{(\star)}{=} \arg \inf_{m,i:X_{m,i}\in\Phi} \left[1-\frac{P_mG_{m,i}}{\|X_{m,i}\|^{\alpha_{m,i}}}\right]^{\omega_m},
\end{align*}
where $I_0$ is defined in the proof of Lemma \ref{Lem:COA} in Appendix \ref{App:ProofCOA} and $(\star)$ is obtained by removing $I_0$ since $I_0$ is the same for all $m,i$. Next, define $X_{*,\mu}$ and $X_{*,\varepsilon}$ as
\begin{align*}
X_{*,\mu}&\defn\arg\inf_{m,i:X_{m,i}\in\Phi\setminus\Phi_M}\left(1-\frac{P_mG_{m,i}}{\|X_{m,i}\|^{\alpha_{m,i}}}\right)^{W_{\mu}}\\
&=\arg\sup_{m,i:X_{m,i}\in\Phi\setminus\Phi_M}\left(\frac{P_mG_{m,i}}{\|X_{m,i}\|^{\alpha_{m,i}}}\right)^{W_{\mu}}
\end{align*}
and 
\begin{align*}
X_{*,\varepsilon}&\defn\arg\inf_{M,i:X_{M,i}\in\Phi_M}\left(1-\frac{P_MG_{M,i}}{\|X_{M,i}\|^{\alpha_{M,i}}}\right)^{W_{\varepsilon}}\\
&=\arg\sup_{M,i:X_{M,i}\in\Phi_M}\left(\frac{P_MG_{M,i}}{\|X_{M,i}\|^{\alpha_{M,i}}}\right)^{W_{\varepsilon}}.
\end{align*}
Note that $X_{*,\mu}$ ($X_{*,\varepsilon}$) represents the UHF (mmWave) BS that provides the typical user with the maximum link rate so that it is exactly the UHF (mmWave) BS that provides the user with the strongest mean received power. Also, we know $X_*\in\{X_{*,\mu},X_{*,\varepsilon}\}$ that can be expressed as
\begin{align}
X_* =\arg\max_{X_{*,\mu},X_{*,\varepsilon}}\left\{\left(\frac{P_mG_{*,\mu}}{\|X_{*,\mu}\|^{\alpha_*}}\right)^{W_{\mu}},\left(\frac{P_MG_M}{\|X_{*,\varepsilon}\|^{\alpha_{M}}}\right)^{W_{\varepsilon}}\right\}. \label{Eqn:OptBS}
%&=\arg\min_{X_{*,\mu},X_{*,\varepsilon}}\left\{\left(1-\frac{P_{*,\mu}G_{*,\mu}}{I_{0,\mu}\|X_{*,\mu}\|^{\alpha_{*,\mu}}}\right),\left(1-\frac{P_MG_M}{I_{0,\varepsilon}\|X_{*,\varepsilon}\|^{\alpha_{M}}}\right)^{\frac{W_{\varepsilon}}{^{W_{\mu}}}}\right\}.
\end{align}
This expression indicates $\Psi_{m,i}=(\frac{P_mG_{m,i}}{\|X_{m,i}\|^{\alpha_{m,i}}})^{\omega_{m}}$ since $\omega_{m}=W_{\mu}\mathds{1}(m\neq M)+W_{\varepsilon}\mathds{1}(m=M)$. Furthermore, \eqref{Eqn:OptBS} can be equivalently written as
\begin{align*}
X_* =\arg\max_{X_{*,\mu},X_{*,\varepsilon}}\left\{\frac{P_mG_{*,\mu}}{\|X_{*,\mu}\|^{\alpha_*}},\left(\frac{P_MG_M}{\|X_{*,\varepsilon}\|^{\alpha_{M}}}\right)^{\frac{W_{\varepsilon}}{W_{\mu}}}\right\},
\end{align*}
which indicates $\Psi_{m,i}(\|X_{m,i}\|)=\frac{P_mG_{m,i}}{\|X_{m,i}\|^{\alpha_{m,i}}}$ whenever $W_{\mu}=W_{\varepsilon}$. This completes the proof.

\bibliographystyle{ieeetran}
\bibliography{IEEEabrv,Ref_mmWaveHetNet}

%\begin{IEEEbiography}[{\includegraphics[width=1in,height=1.25in,clip,keepaspectratio]{ChunHungLiu}}]{Chun-Hung Liu}(M'08 -- SM'16) received the B.S. degree in Mechanical Engineering and the M.S. degree in Electrical Engineering from National Taiwan University, Taipei, Taiwan, the M.S. degree in Mechanical Engineering from Massachusetts Institute of Technology, Cambridge, MA, USA, and the Ph.D. degree in Electrical and Computer Engineering from The University of Texas at Austin, TX, USA. He is currently an Assistant Professor with the Department of Electrical and Computer Engineering at Mississippi State University, Starkville MS, USA. He was with the University of Michigan, Ann Arbor MI, USA and National Chiao Tung University, Hsinchu, Taiwan. His research interests include wireless communication, information theory, stochastic geometry and random graph, machine learning, networked control and optimization theory. He was a recipient of the Best Paper Award from IEEE Globecom in 2014, a recipient of the Excellent Young Researcher Award from the Ministry of Science and Technology of Taiwan in 2015, a recipient of the Outstanding Advisor Award from the Taiwan Institute of Electrical and Electronic Engineering in 2016.
%\end{IEEEbiography}

\end{document}